\documentclass[11pt]{article}

\usepackage[T1]{fontenc}
\usepackage[utf8]{inputenc}
\usepackage{newtxtext,newtxmath}
\usepackage{microtype}
\usepackage[letterpaper,margin=1in]{geometry}
\usepackage{amsmath,mathtools,bm}
\usepackage{booktabs,tabularx,array,multirow}
\usepackage{siunitx}
\usepackage{graphicx}
\usepackage{xcolor}
\usepackage[font=small,labelfont=bf,labelsep=period]{caption}
\usepackage{cite}
\usepackage{url}
\usepackage[hidelinks]{hyperref}
\usepackage[capitalise,noabbrev]{cleveref}

\newcommand{\DataDOI}{10.5281/zenodo.23608384}

\hypersetup{
  pdftitle={The Analytical Investigation of Inductor-Coupled Mixers},
  pdfauthor={Arash Ghasemi},
  pdfkeywords={balanced mixer, closed-form solution, diode ring, Gilbert cell, inductive coupling, Lambert W function, r-Lambert function}
}

\graphicspath{{figures/}}
\newcommand{\CircuitWidth}{0.8\textwidth}
\newcommand{\PlotWidth}{0.62\textwidth}

\newcommand{\pospart}[1]{\left[#1\right]_{+}}

\newcommand{\rLambert}[2]{\mathcal W_{#1}\!\left(#2\right)}
\newcommand{\asinh}{\operatorname{asinh}}
\newcommand{\artanh}{\operatorname{artanh}}
\newcommand{\arcosh}{\operatorname{arcosh}}
\newcommand{\sgn}{\operatorname{sgn}}
\newcommand{\dif}{\mathop{}\!\mathrm{d}}

\title{The Analytical Investigation of Inductor-Coupled Mixers}
\author{Arash Ghasemi\thanks{Email: \href{mailto:syn535@tennessee.edu}{syn535@tennessee.edu}}\\[0.6ex]
  {\small Sumer Analog LLC, Knoxville, TN, USA}\\
  {\small University of Tennessee System, USA}}
\date{\today}

\begin{document}
\maketitle

\begin{abstract}
Inductor- and transformer-coupled balanced mixers are usually analyzed with
switching approximations, conversion matrices, harmonic balance, or circuit
simulation.  This paper instead derives their loaded quasistatic equations
from the device laws, keeping the transformer ratios, reflected load, bias,
and the connections that set the IF common mode.  The
functions needed to solve these equations are surprisingly varied.  A
Shockley single-balanced pair is solved by the principal Lambert function
\(W_0\) under a quantified one-diode-dominance condition.  The double-balanced
ring has a unique loaded solution that reduces to an exact
inverse-hyperbolic-tangent form only for an open IF port.  Vacuum diodes and
triodes lead to Cardano cubics and structured sextics, whereas depletion
JFET/MESFET crossbars need only square roots.  Bilateral MOS channels give
piecewise quadratics in strong inversion, or a Ferrari quartic when the IF
winding floats.  In weak inversion they give an exact \(W_0\), the real
\(r\)-Lambert function, or the inverse of \(\sinh x+\rho x\), depending on
topology and IF connection.  In matched BJT and HBT Gilbert cells, equal
emitter resistors turn the hyperbolic-tangent current steering into an exact
\(r\)-Lambert function.  Fourier projection brings in Beta/Gamma, modified
Bessel, and complete elliptic functions.  We checked the equations against 37
LTspice simulations using seven independent Python programs.  In the 12
primary comparisons, which implement the assumed device laws exactly, every
RMS waveform error was below \(0.017\%\) of the analytical RMS value;
built-in and vendor device models were examined separately.  Two design
examples show how the closed forms size a thermionic IF load and a
Gilbert-cell tail current without parameter sweeps.
\end{abstract}


\pdfbookmark[1]{Nomenclature}{nomenclature}
\section*{Nomenclature}

Symbols are grouped by topic, and a symbol that is reused in a later group
is redefined there.  Signs and polarities follow the schematics, and standard
mathematical operators have their usual meanings.

\begingroup
\small
\raggedright
\setlength{\parskip}{1.2pt}
\newcommand{\mixernom}[2]{\par\noindent\hangindent=1em\hangafter=1\ensuremath{#1}\enspace #2\par}

\subsection*{Conventions and functions}

\mixernom{\widehat{x}}{Peak amplitude of \(x\).}
\mixernom{\dot{x},f'(x),\partial,\dif}{Time, argument, and partial derivatives; differential.}
\mixernom{\pospart{x},|\mathcal A|}{Positive part \(\max(x,0)\); set cardinality.}
\mixernom{j,k,\sigma}{Indices; \(\sigma=\pm\) selects the control phase.}
\mixernom{S,D,\chi}{Single-/double-balanced configurations; \(\chi\in\{S,D\}\).}
\mixernom{W_0,W_{-1}}{Principal/lower real Lambert branches: \(W e^W=z\).}
\mixernom{\mathcal W_r(a)}{Solution \(z\) of \(ze^z+rz=a\); \(r\) is a function parameter.}
\mixernom{\sinh x+\rho x}{Monotone map; load coefficient \(\rho\ge0\), denoted \(\rho_c\) in weak MOS.}
\mixernom{\Gamma,I_1(x)}{Gamma function; modified Bessel function, first kind, order one.}
\mixernom{K(m),E(m)}{Complete elliptic integrals, first/second kind; modulus \(m\).}
\mixernom{O(x^k)}{Big-O asymptotic bound.}

\subsection*{Signals, transformers, and loads}

\mixernom{t,\omega,\omega_R,\omega_L,\theta}{Time; angular frequencies (RF: \(R\), LO: \(L\)); phase.}
\mixernom{v_s,\widehat V_s,u}{Source voltage, its peak, and primary magnetizing voltage.}
\mixernom{v_{\rm RF},v_{\rm LO}}{External RF/LO drives; peaks \(\widehat V_{\rm RF},\widehat V_{\rm LO}\).}
\mixernom{r,\ell}{Core RF/LO half-swings; normalized in the Shockley ring.}
\mixernom{s}{LO sign \(\sgn\ell\).}
\mixernom{v,v_0(\ell)}{Core output and zero-RF value; crossbar \(v=(p-n)/2\).}
\mixernom{v_{\rm IF}}{Extracted IF waveform; instantaneous output in the Shockley ring.}
\mixernom{v_{{\rm IF},o},\widehat V_{\rm IF}}{External IF-port voltage; IF peak amplitude.}
\mixernom{x_\pm,v_1,v_2}{Single-balanced endpoints: \(x_+=v_1=\ell+r\), \(x_-=v_2=\ell-r\).}
\mixernom{v_3,v_4,v_5,v_6}{Ring nodes: \(-\ell,v-r,\ell,v+r\).}
\mixernom{a,b,p,n}{RF endpoints \(a,b\); IF endpoints \(p,n\).}
\mixernom{c}{IF common mode: \((p+n)/2\), or \((P+N)/2\).}
\mixernom{V_B,V_G}{RF center-tap bias; LO gate/grid common-mode bias.}
\mixernom{g,g_\pm}{Control voltage; complementary gate/grid/base drives.}
\mixernom{I_p,I_n,I_+,I_-}{IF-node injections; oriented branch currents in \eqref{eq:generic-kcl-paper}.}
\mixernom{F,\Phi,\Phi_\pm,\Phi_1,\ldots,\Phi_4}{Loaded KCL residual; nonlinear branch-current laws.}
\mixernom{L_p,R_s}{Primary magnetizing inductance; source resistance.}
\mixernom{T_{\rm RF},T_{\rm LO},K}{RF/LO transformers; magnetic coupling coefficient.}
\mixernom{N_{R,p},N_{R,s}}{RF primary/secondary turns; also \(N_{\mathrm{RF},p},N_{\mathrm{RF},s}\).}
\mixernom{N_{L,p},N_{L,s}}{LO primary/secondary turns; also \(N_{\mathrm{LO},p},N_{\mathrm{LO},s}\).}
\mixernom{N_{I,p},N_{I,s}}{IF primary/secondary turns; \(N/2\): half a center-tapped winding.}
\mixernom{n_R,n_L,n_I}{RF/LO/IF secondary-to-primary turns ratios; \(n_{\rm RF}=n_R\), \(n_{\rm LO}=n_L\).}
\mixernom{R_o,R_\Delta}{External IF load; reflected differential load \(R_o/n_I^2\).}
\mixernom{R_L}{Diode/MOS load; full differential load for a double MOS bridge.}
\mixernom{R_S,R_{\rm eff}}{Single FET/triode load; effective scalar-KCL load.}
\mixernom{\mathcal P_\omega}{Fourier projection at \(\omega\), including the stated IF transfer.}
\mixernom{g(\ell),G_1}{Small-RF transfer coefficient; its first LO cosine coefficient.}
\mixernom{G_{1,\chi},\widehat V_{{\rm IF},o,\chi}}{First LO coefficient; external first-sideband peak for \(\chi\).}

\subsection*{Shockley diodes}

\mixernom{I_s,V_D}{Saturation current; ideality-scaled thermal voltage.}
\mixernom{a,b,x}{Single-pair normalized voltages: \(r/V_D,\ell/V_D,v/V_D\).}
\mixernom{\xi}{Inverse load-current scale: \(V_D/(I_sR_L)\).}
\mixernom{v_W}{One-diode-dominance Lambert output.}
\mixernom{\epsilon}{Allowed discarded/retained current ratio.}
\mixernom{z_0,w_0}{Zero-RF Lambert argument \((I_sR_L/V_D)e^{|\ell|/V_D}\); \(w_0=W_0(z_0)\).}
\mixernom{u,r,y}{Ring-normalized LO, RF, output: \(n_Lv_{\rm LO}/(2V_D)\), \(n_Rv_{\rm RF}/(2V_D)\), \(v_{\rm IF}/V_D\).}
\mixernom{z,c}{Unknown/right-hand side in \(ze^z=c\).}

\subsection*{Vacuum diodes}

\mixernom{\mathcal P,V_c}{Perveance (\(\mathrm{A\,V^{-3/2}}\)); conduction-knee voltage.}
\mixernom{\Phi_T(z)}{Current at anode-to-cathode voltage \(z\).}
\mixernom{B,V}{Single-path drive \(|\ell|+r-V_c\); output magnitude, \(v=sV\).}
\mixernom{\gamma,w,\eta}{Load factor \(R_L\mathcal P\); Cardano variable \(\gamma\sqrt{B-V}\); cubic forcing \(\gamma^2B\).}
\mixernom{i_{jk}}{Current from ring node \(j\) to \(k\).}
\mixernom{A,z,c}{Ring variables: \(|\ell|-V_c\), \(sv\), \((\mathcal PR_L)^{-1}\).}
\mixernom{g_T(\ell)}{Ring small-RF transfer coefficient.}
\mixernom{\zeta}{Conduction parameter \(3\mathcal PR_L\sqrt{\widehat\ell}\), for \(V_c=0\).}
\mixernom{C_\Gamma,\eta_{\rm min}}{Load-sizing constant \(1.19814\); minimum required fraction of the ideal first-sideband amplitude.}
\mixernom{C_{AK}}{Anode-to-cathode capacitance.}

\subsection*{Depletion FETs}

\mixernom{V_P,P_J,I_{\rm DSS}}{Pinch-off voltage; magnitude \(P_J=-V_P\); zero-gate-bias saturation current.}
\mixernom{\beta_J}{Square-law coefficient \(I_{\rm DSS}/P_J^2\).}
\mixernom{J_J(x,y;g)}{Channel current from \(x\) to \(y\), controlled by \(g\).}
\mixernom{u_\pm,U,u_H}{Control offsets: \(P_J+V_G\pm\ell\), \(P_J+V_G\), \(U+|\ell|\).}
\mixernom{\rho_J,\rho_{J,S}}{Load coefficients \((\beta_JR_\Delta)^{-1}\), \((\beta_JR_S)^{-1}\) (volts).}
\mixernom{h(u,x)}{Endpoint difference \([u+x]_+^2-[u-x]_+^2\); also used for MOS.}
\mixernom{\mathcal A,m}{Active output-term set; size \(m=|\mathcal A|\).}
\mixernom{S,C,D}{Single-pair sums: \(\sum_{\mathcal A}u_\sigma\), \(\sum_{\mathcal A}u_\sigma^2\), and input-side positive-part squares.}
\mixernom{g_{J,S},g_{J,D}}{Single-/double-balanced small-RF transfer coefficients.}
\mixernom{Q_1,\ldots,Q_4}{Crossbar FETs.}

\subsection*{Vacuum triodes}

\mixernom{B^+,\mathcal P_3,\mu}{Plate supply \(V_B\); perveance; amplification factor.}
\mixernom{P,K,G}{Plate/cathode/grid potentials; also \(V_p,V_k,V_g\).}
\mixernom{V_{GK},V_{PK}}{Grid-to-cathode and plate-to-cathode voltages.}
\mixernom{I_3,\Phi_3(E),E}{Plate current, three-halves law, and control voltage \(V_{GK}+V_{PK}/\mu-V_c\).}
\mixernom{A,\kappa}{Control bias \(V_G+V_B/\mu-V_c\); cathode factor \(1+1/\mu\).}
\mixernom{E_{pa},E_{na},E_{pb},E_{nb}}{Branch controls; suffixes identify cathode then plate.}
\mixernom{G_\pm}{Conductances at \(r=v=0\): \((3\mathcal P_3/2)[A\pm\ell]_+^{1/2}\).}
\mixernom{g_{3,D}}{Double-balanced small-RF transfer coefficient.}
\mixernom{z,H,\delta}{Selected-diagonal variables: \(sv\), \(A+|\ell|\), \(r/\mu-\kappa z\).}
\mixernom{D}{Radical/sextic load term \(4z/(\mathcal P_3R_\Delta)\).}
\mixernom{B_s,\gamma,w}{Single-triode drive \(A+|\ell|+sr/\mu\); load factor \(\kappa\mathcal P_3R_S\); Cardano variable \(\gamma\sqrt{B_s-\kappa v}\).}
\mixernom{T_1,\ldots,T_4,H_1,\ldots,H_4}{Triodes and their isolated heater connections.}

\subsection*{Strong-inversion MOS}

\mixernom{\beta,V_{T0}}{Square-law coefficient (\(\mathrm{A/V^2}\)); fixed threshold voltage.}
\mixernom{\Xi_g(x)}{Endpoint-current potential \(\beta[g-V_{T0}-x]_+^2/2\).}
\mixernom{J_\square(a,b;g)}{Square-law channel current from \(a\) to \(b\), controlled by \(g\).}
\mixernom{u_\pm,U}{Gate overdrives: \(g_\pm-V_{T0}\), \(V_G-V_{T0}\).}
\mixernom{A}{Single-pair quadratic coefficient \(2U+1/(\beta R_L)\).}
\mixernom{C_G,C_I,\Delta_B}{Gate/IF common-mode references; RF-minus-IF common-mode voltage.}
\mixernom{\bar u_\pm}{IF-referenced overdrives \(C_G-C_I-V_{T0}\pm\ell\).}
\mixernom{R_h,P,N}{IF endpoint-to-reference resistance; relative endpoint voltages.}
\mixernom{g_\square(\ell)}{Double-core small-RF transfer coefficient.}
\mixernom{q,a}{Floating-core overdrive \(U-c\); load parameter \((\beta R_L)^{-1}\).}
\mixernom{\kappa_s,q_H,q_L}{Global single-pair variables: \(\beta R_L\), \(U+|\ell|\), \(U-|\ell|\).}
\mixernom{\Delta,D_s}{Overdrive separation \(2|\ell|\); half-sum of squared RF-side positive parts.}
\mixernom{G_H,G_L}{Upper/lower conduction-state tests in Appendix~\ref{app:mos-global}.}
\mixernom{E}{Two-active-output coefficient \(U+\kappa_s(D_s-\ell^2)\).}

\subsection*{Weak-inversion MOS}

\mixernom{J_w(a,b;g)}{Weak-inversion channel current from \(a\) to \(b\).}
\mixernom{I_0,U_T,n}{Current scale; thermal voltage; subthreshold slope factor.}
\mixernom{A_\pm,A_0}{Current prefactors \(I_0e^{(g_\pm-V_{T0})/(nU_T)}\); zero-LO value \(A_0\).}
\mixernom{\rho,x}{Normalized RF/IF half-voltages: \(r/U_T,v/U_T\).}
\mixernom{C,D}{Single-pair load/input coefficients in \(x=Ce^{-x}-D\).}
\mixernom{V_{P,j},z}{Device pinch-off potential; channel-endpoint voltage.}
\mixernom{\rho_c,q_c}{Normalized load/drive in \(\sinh x+\rho_cx=q_c\).}
\mixernom{\xi,y}{Floating common-mode/half-differential voltages: \(c/U_T,v/U_T\).}
\mixernom{s_\pm}{Normalized channel strengths \(R_LA_\pm/(2U_T)\).}
\mixernom{d_+,d_-}{RF coefficients: \(s_+e^{-\rho}+s_-e^\rho\), \(s_+e^\rho+s_-e^{-\rho}\).}
\mixernom{z}{Floating \(r\)-Lambert unknown \(2(y+d_+)\).}
\mixernom{g_w,\chi}{Double-core small-RF transfer; load parameter \(R_LA_0/U_T\).}
\mixernom{\lambda,\Lambda}{Normalized LO \(\ell/(nU_T)\) and peak: \(\lambda=\Lambda\cos\theta\).}
\mixernom{q_B}{Bessel-series parameter \((1-\sqrt{1-\chi^2})/\chi\).}

\subsection*{Active MOS pair}

\mixernom{I_T,i_1,i_2}{Tail current; drain currents.}
\mixernom{v_d,z}{Differential gate voltage; normalized drive \(v_d/[2\sqrt{I_T/\beta}]\).}
\mixernom{m,S_1(m)}{Peak drive in \(z=m\cos\theta\); first cosine coefficient of normalized current difference.}

\subsection*{BJT/HBT}

\mixernom{I_C,I_S,V_{BE},U_E}{Collector/saturation currents; base-emitter voltage; exponential voltage scale.}
\mixernom{I_1,I_2,J}{Pair collector currents; available total.}
\mixernom{U_R,U_L}{RF/LO-pair exponential voltage scales.}
\mixernom{I_T,I_{E,T}}{Available collector-current scale; emitter tail-source current.}
\mixernom{J_a,J_b}{RF branch currents; schematic labels \(I_a,I_b\).}
\mixernom{i_\Delta,i_{\Delta,S},i_{\Delta,D}}{Differential output current; single-/double-balanced forms.}
\mixernom{T_1(a)}{First cosine coefficient of \(\tanh(a\cos\theta)\); normalized drive peak \(a\).}
\mixernom{a_R,a_L}{RF/LO drives: \(\widehat r/U_R,\widehat\ell/U_L\) for sidebands; \(r/U_R,\ell/U_L\) for degenerated waveforms.}
\mixernom{\widehat V_{{\rm IF},S}^{(1,1)},\widehat V_{{\rm IF},D}^{(1,1)}}{First sum/difference sideband peaks; \((1,1)\): fundamental RF/LO mixing.}
\mixernom{J_E,I_{E1},I_{E2},R_E}{Total/individual pair emitter currents; resistance per emitter.}
\mixernom{v_d,m}{Differential base voltage; normalized current difference \((I_{E1}-I_{E2})/J_E\).}
\mixernom{a,b,w}{Pair drive \(v_d/(2U_E)\); degeneration \(R_EJ_E/(2U_E)\); substitution \(2b(m+1)\).}
\mixernom{\mathscr S(a,b)}{Normalized current division solving \(m=\tanh(a-bm)\).}
\mixernom{b_R,x_E}{RF degeneration \(R_EJ_E/(2U_R)\); current division \(\mathscr S(a_R,b_R)\).}
\mixernom{R_{E,L},s_a,s_b}{LO emitter resistance; upper-pair normalized current divisions.}
\mixernom{\alpha,\alpha_R,\alpha_L}{Collector/emitter current ratios; RF/LO-stage values.}
\mixernom{V_{B,R},V_{B,L},r_\pm}{RF/LO base biases; complementary RF drives.}
\mixernom{V_{CC},V_{EE},R_D}{Collector/tail-return supplies; dummy collector load.}
\mixernom{Q_{R\pm},Q_{L\pm},Q_{D\pm}}{RF, active LO, and dummy transistor pairs.}
\mixernom{Q_{A\pm},Q_{B\pm}}{Double-balanced upper transistor pairs.}

\subsection*{Coupled inductors}

\mixernom{\bm i_w,\bm v_n,\bm v_s}{Winding-current, nonlinear-node-voltage, source-voltage vectors.}
\mixernom{\bm L,\bm R_w}{Self/mutual-inductance and winding-resistance matrices.}
\mixernom{\bm A,\bm B}{Winding-incidence and source-connection matrices.}
\mixernom{\bm j_{\rm nl},\bm G,\bm q}{Nonlinear current injections; nodal conductance; stored node charge.}
\mixernom{(\cdot)^{\mathsf T},\bm0}{Transpose; zero vector. Bold symbols denote vectors/matrices.}

\subsection*{Validation and abbreviations}

\mixernom{N,k,v_{{\rm LT},k},v_{{\rm eq},k}}{Sample count/index; simulated/analytical voltage samples.}
\mixernom{\mathrm{NRMSE}}{RMS waveform error / analytical RMS, expressed as a percentage.}
\mixernom{\mathrm{RF},\mathrm{LO},\mathrm{IF},\mathrm{CT}}{Radio frequency; local oscillator; intermediate frequency; center tap.}
\mixernom{\mathrm{FET},\mathrm{JFET},\mathrm{MESFET}}{Field-effect transistor; junction FET; metal-semiconductor FET.}
\mixernom{\mathrm{MOS},\mathrm{MOSFET},\mathrm{HEMT}}{Metal-oxide-semiconductor; MOS FET; high-electron-mobility transistor.}
\mixernom{\mathrm{BJT},\mathrm{HBT},\mathrm{NPN}}{Bipolar junction transistor; heterojunction bipolar transistor; n-p-n structure.}
\mixernom{\mathrm{GaAs},\mathrm{SiGe}}{Gallium arsenide; silicon-germanium.}
\mixernom{\mathrm{KCL},\mathrm{FFT},\mathrm{RMS}}{Kirchhoff's current law; fast Fourier transform; root mean square.}
\mixernom{\mathrm{CAD},\mathrm{EM},\mathrm{WDF}}{Computer-aided design; electromagnetic; wave-digital filter.}
\mixernom{\mathrm{EKV},\mathrm{SPICE}}{Enz--Krummenacher--Vittoz; Simulation Program with Integrated Circuit Emphasis.}

\endgroup
 \section{Introduction and Prior Work}
\label{sec:introduction}

In a balanced mixer the nonlinear devices perform the frequency conversion,
but the circuit around them decides what actually reaches the IF port.  The
transformer turns ratios set the voltage swings at the device terminals, the
winding polarities determine which products cancel, the IF load closes the
nonlinear circuit equation, and the center taps fix the common-mode voltage.
These details are what distinguish a single-balanced pair from a
double-balanced ring, even when both are built from the same device.  Modern
broadband designs still put considerable effort into balun amplitude and
phase balance and into IF filtering
\cite{Wang2023Broadband}, and commercial passive-FET mixers continue to use a
four-device crossbar surrounded by three coupled ports
\cite{MiniCircuits2015}.

\subsection{Prior work}

The theory of balanced rectifier modulators driven by a strong carrier is
much older than modern RF CAD.  Caruthers analyzed carrier-controlled
copper-oxide modulators \cite{Caruthers1939}, and Belevitch developed a linear
switched-resistance theory of bridge and ring modulators
\cite{Belevitch1948}.  Saleh later formulated a network theory of resistive
mixers as periodically time-varying circuits \cite{Saleh1971}.  Subsequent
work used response coefficients and conversion matrices to analyze the
response, conversion loss, and noise of double-balanced mixers
\cite{MaiuzzoCameron1979,HeldKerr1978}, and identified finite commutation and
imperfect diode switching as sources of intermodulation \cite{Walker1976}.
Most of this work starts from a switched or linearized device description.
Here we instead solve the nonlinear circuit equations with the IF load in
place, and only then extract the wanted frequency components from the
resulting waveform.

Each of the standard analysis tools answers a different question.  Taylor
and Volterra expansions expose low-order intermodulation, conversion matrices
and harmonic balance keep track of many spectral components at once, and
transient SPICE simulation includes compact device models and parasitics.  A
recent nodal Volterra analysis, for example, predicts third-order distortion
in balanced, double-balanced, and triple-balanced diode mixers to within
\(3\,\mathrm{dB}\) of simulation \cite{Golovan2024}.  These methods are
indispensable once a topology has been chosen.  What they usually do not give
the designer is the loaded instantaneous waveform as a compact function of
the device and transformer parameters.

Closed-form inversions have appeared before, but in narrower settings.  A
diode that obeys Shockley's exponential law \cite{Shockley1949} in series
with a resistor can be solved exactly with the Lambert \(W\) function
\cite{Banwell2000,Corless1996}.  For the diode ring, Parker used a static
polynomial reduction selected by the carrier \cite{Parker2011}, whereas
Bernardini \emph{et al.} derived Lambert-\(W\) wave-digital elements for a
simplified ring model \cite{Bernardini2016}.  Lambert \(W\) also appears in
analytical MOS electrostatics \cite{OrtizConde2003}.  To our knowledge,
however, no previous work has established which loaded balanced-mixer
equations reduce to the ordinary \(W\) function, which ones require other
inverse functions, and how the answer depends on the device law, the load,
and the bias connections.

The devices themselves have an equally long history.  Frequency conversion
with thermionic diodes had been characterized experimentally by 1939
\cite{Gainsborough1939}, and the space-charge law that governs these diodes
is due to Child and Langmuir \cite{Child1911,Langmuir1913}.  Driving FET
switching elements in antiphase from a transformer was patented in 1968
\cite{Squires1968}.  Pucel \emph{et al.} reported measurements and models of
GaAs MESFET mixers in the X band, roughly \(8\) to \(12\,\mathrm{GHz}\)
\cite{Pucel1976}.  For MOS channels, the long-channel square-law model
\cite{ShichmanHodges1968} and the all-region EKV model \cite{Enz1995} provide
two useful asymptotic laws.  Among active mixers, the double-balanced Gilbert
mixer and its hyperbolic-tangent steering law rest on the translinear core of
Gilbert's matched bipolar four-quadrant multiplier \cite{Gilbert1968}.  The
circuits and the special functions used here each have their own history;
what is new is combining them in a way that accounts for the circuit
topology.

The transformers and baluns have a design literature of their own.
Ruthroff's broadband transformers \cite{Ruthroff1959} and monolithic-balun
mixers \cite{Chen1991} address bandwidth, balance, and matching rather than
the pointwise solution of the nonlinear equations.  Work on resistive FET
mixers ranges from symmetric MOS mixers \cite{Rafuse1968} and
low-intermodulation MESFET designs \cite{Maas1987,DeFlaviisMaas1995} to a
simplified analysis, aimed at hand calculation, based on time-varying
reflection coefficients \cite{Yhland2007}.  Wideband SiGe-HBT Gilbert mixers
have likewise been analyzed with conversion matrices and harmonic balance and
verified by measurement \cite{Johansen2005}.  Our formulas belong to an
earlier stage.  They give first component estimates, under clearly stated
load and bias assumptions, before detailed frequency-domain and
implementation work begins.

\Cref{tab:literature-paper} summarizes how the most closely related studies
compare with the present analysis.

\begin{table*}[!t]
\caption{Representative prior work and its relation to the present
analysis.}
\label{tab:literature-paper}
\centering
\footnotesize
\setlength{\tabcolsep}{4.5pt}
\begin{tabularx}{\textwidth}{@{}p{0.18\textwidth}p{0.23\textwidth}p{0.27\textwidth}X@{}}
\toprule
Reference & Device/topology & Principal method or result & Relation to this work \\
\midrule
Gilbert \cite{Gilbert1968} & matched bipolar four-quadrant multiplier & exact current steering in an exponential pair & Provides the canonical \(\tanh\) core.  We add the transformer coordinates, the IF projection, and exact \(r\)-Lambert steering with emitter degeneration. \\
Pucel \emph{et al.} \cite{Pucel1976} & microwave GaAs MESFET mixers & device and mixer performance analysis with experiment & Motivates the treatment of FETs other than MOS.  Our endpoint-law crossbar gives loaded roots that are exact within stated gate-bias regions. \\
Banwell and Jayakumar \cite{Banwell2000} & diode with series resistance & exact Lambert-\(W\) current & Establishes the exponential-plus-resistor solution, but for a single diode rather than a balanced transformer-coupled pair. \\
Parker \cite{Parker2011} and Bernardini \emph{et al.} \cite{Bernardini2016} & diode ring-modulator models & carrier-selected polynomial reduction; exact Lambert-\(W\) WDF elements & Already apply Lambert \(W\) to ring models, so we make no general priority claim.  Our \(W_0\) result is tied to a stated one-diode-dominance condition and a specific loaded topology. \\
Mini-Circuits \cite{MiniCircuits2015} & three-balun passive FET crossbar & practical switching topology and measured product family & Provides the connection graph.  We do not guess the unpublished channel law, and we keep the RF center-tap bias as a symbol. \\
Golovan \cite{Golovan2024} & balanced, double-, and triple-balanced diode mixers & generalized nodal/Volterra distortion analysis & A complementary frequency-domain perturbation method.  We instead solve the loaded instantaneous circuit equation. \\
Wang \emph{et al.} \cite{Wang2023Broadband} & 1--6 GHz GaAs double-balanced mixer & balun/filter design, EM/circuit simulation, and measurement & Illustrates a modern workflow spanning analysis, simulation, fabrication, and measurement.  Our formulas supply a sizing step that comes before it. \\
\bottomrule
\end{tabularx}
\end{table*}

\subsection{Contributions and scope}

The paper makes five contributions.

\begin{enumerate}
  \item We define a single coordinate system for inductively coupled mixers
  that keeps the turns ratios, the reflected load, the winding polarities, the
  bias, and the IF common mode.  It separates the quasistatic nonlinear
  equation from the finite-bandwidth differential equations of the
  transformers.
  \item We derive the loaded single- and double-balanced equations for
  Shockley and thermionic diodes, bilateral MOS channels in strong and weak
  inversion, depletion JFET/MESFET and vacuum-triode crossbars, and matched
  BJT/HBT half-Gilbert and Gilbert cells.
  \item For each device law and circuit connection, we identify the function
  that solves the equation: the ordinary \(W_0\), the real \(r\)-Lambert
  function \cite{MezoBaricz2017}, a cubic solvable by Cardano's formula, a
  quartic solvable by Ferrari's method, a structured sextic, the unique
  inverse of \(x\mapsto\sinh x+\rho x\), or, when none of these applies, the
  root of a monotone scalar equation.  One consequence is that changing only
  the way the IF common-mode voltage is set can change the required function,
  even when the MOS devices are the same.
  \item We validate the equations against 37 LTspice simulations, analyzed by
  seven independent Python programs.  Simulations that implement exactly the
  device laws used in the derivations are reported separately from those that
  use LTspice's built-in or manufacturer-supplied device models.
  \item We show how the formulas support hand calculation in two design
  problems, sizing the IF load of a thermionic mixer and the tail current of a
  BJT Gilbert cell, and we check each estimate against the full scalar
  equation and LTspice.
\end{enumerate}

The analysis is deliberately confined to the front end of the design process
and is built directly on the device laws.  Leakage, breakdown, charge
storage, winding loss, imperfect coupling, and frequency-dependent
terminations are left to compact-model and electromagnetic simulation, which
follow the analytical estimate.  All results reported here come from analysis
and simulation; none are hardware measurements.
 \section{Inductive-Coupling Framework}
\label{sec:framework}

\subsection{Finite inductance and ideal transformer voltage ratios}

Let a source with resistance \(R_s\) drive a primary inductance \(L_p\), and
let \(u\) be the voltage across the idealized magnetizing branch.  Before any secondary loading is reflected to
the primary, \(u\) satisfies
\begin{equation}
 \dot u+\frac{R_s}{L_p}u=\dot v_s.
 \label{eq:primary-ode-paper}
\end{equation}
For a sinusoidal source \(v_s=\widehat V_s\cos\omega t\), the steady-state
magnitude ratio is that of a first-order high-pass filter with corner
frequency \(R_s/L_p\):
\begin{equation}
 \left|\frac{u}{v_s}\right|
 =\frac{\omega}{\sqrt{\omega^2+(R_s/L_p)^2}}.
 \label{eq:primary-transfer-paper}
\end{equation}
The swing on each half of the RF secondary is then \(r=n_Ru/2\), where
\(n_R=N_{R,s}/N_{R,p}\), and the LO transformer gives
\(\ell=n_Lv_{\rm LO}/2\) in the same way.  For the IF transformer we define
\(n_I=N_{I,s}/N_{I,p}\).  An external differential load \(R_o\) then appears
at the core as \(R_\Delta=R_o/n_I^2\), and the bridge half-voltage \(v\)
appears at the IF port as \(v_{{\rm IF},o}=2n_Iv\).

Equations \eqref{eq:primary-ode-paper} and
\eqref{eq:primary-transfer-paper} are where the physical inductance enters a
hand calculation: they show whether the magnetizing inductance is large
enough at the operating frequency.  Everything that follows assumes ideal
transformer voltage ratios and known drives \(r(t)\) and \(\ell(t)\).  Once
winding currents, mutual inductance, source impedance, and parasitic charge
are included, the formulas no longer solve the whole problem; they become the
nonlinear relations inside the differential-algebraic system of
Appendix~\ref{app:dae-paper}.

\Cref{fig:generic-topologies-paper} shows the single- and double-balanced
templates used throughout the paper.

\begin{figure}[!t]
 \centering
 \includegraphics[width=\CircuitWidth]{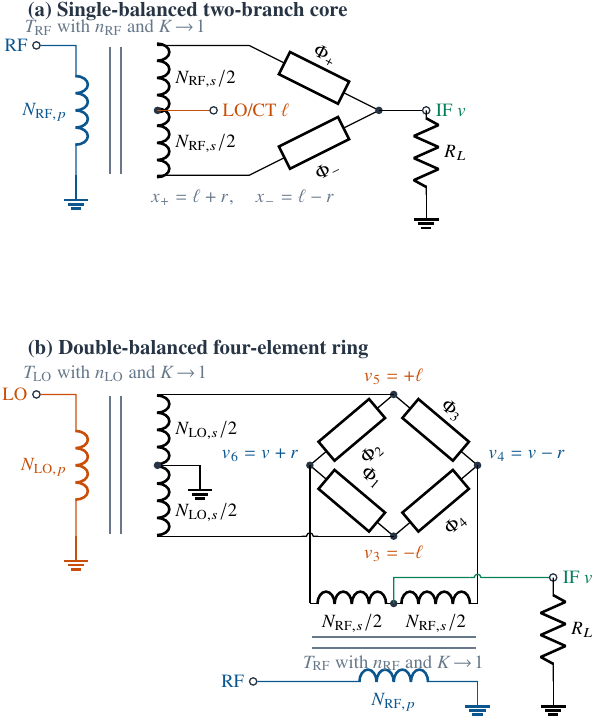}
 \caption{Parametric inductively coupled templates.  The single-balanced core
 uses two nonlinear branches; the double-balanced core uses a four-element
 ring and a center-tapped IF winding.  The symbol \(\Phi\) denotes either a
 two-terminal device law or a fully specified controlled channel;
 three-terminal devices need the control networks shown in later figures.
 The transformer dots set the signs of \(r\), \(\ell\), and the external IF
 voltage.}
 \label{fig:generic-topologies-paper}
\end{figure}

\subsection{Loaded scalar equation and IF projection}

With ideal voltage ratios, the nonlinear core is described by one or two
algebraic unknowns.  A single-balanced pair has the endpoint drives
\begin{equation}
 x_+=\ell+r,\qquad x_-=\ell-r
 \label{eq:single-coordinates-paper}
\end{equation}
and a single output \(v\).  In a double-balanced bridge whose IF center tap
is held at a fixed voltage, the IF endpoints are \(p=+v\) and \(n=-v\),
measured from that reference.  If the IF secondary floats instead, a second
unknown appears:
\begin{equation}
 p=c+v,\qquad n=c-v,
 \label{eq:floating-coordinates-paper}
\end{equation}
where the common-mode voltage \(c\) is the average of the two IF endpoint
voltages.  Its value is fixed by the requirement that no net current flows
into the isolated winding.  This one extra unknown is what produces the MOS
quartic and one of the \(r\)-Lambert results later in the paper.

For the passive circuits we orient the nonlinear branch currents so that KCL
at the output reads
\begin{equation}
 F(v;r,\ell)=\frac{v}{R_{\rm eff}}-I_+(v;r,\ell)
             +I_-(v;r,\ell)=0.
 \label{eq:generic-kcl-paper}
\end{equation}
Every device law used below is monotone in the passive direction, so
\(\partial F/\partial v>0\) (or the reverse inequality with the opposite
orientation), and the circuit equation has exactly one physical solution.
When the solution can be written with a named function, that function is
simply a convenient way to compute it.  When no named inverse exists, the root can still be found
quickly and reliably by a bracketing method, such as bisection, on an
interval known to contain it.

\Cref{tab:function-map-paper} collects the solution families obtained for the
different device laws and circuit connections.

\begin{table*}[!t]
\caption{Device laws and the functions that solve the loaded mixer equations
for the stated IF load and common-mode connections.}
\label{tab:function-map-paper}
\centering
\footnotesize
\setlength{\tabcolsep}{4.5pt}
\begin{tabularx}{\textwidth}{@{}p{0.16\textwidth}p{0.29\textwidth}p{0.24\textwidth}X@{}}
\toprule
Device family & Device law (memoryless) & Single-balanced result & Double-balanced result \\
\midrule
Junction diode & \(I_s(e^{v/V_D}-1)\) & unique root of a two-exponential equation; exact \(W_0\) when one diode dominates & unique loaded root; exact inverse-hyperbolic-tangent form only when the IF port is open \\
Thermionic diode & \(\mathcal P[v-V_c]_+^{3/2}\) & Cardano cubic when only one tube conducts & radical equation, which becomes a structured sextic after elimination \\
Depletion JFET/MESFET & symmetric endpoint square law & piecewise quadratic, square roots only & piecewise linear or quadratic, square roots only \\
Vacuum triode & \(\mathcal P_3[V_g+V_p/\mu-(1+1/\mu)V_k-V_c]_+^{3/2}\) & Cardano cubic when only one tube conducts & monotone radical equation, which becomes a structured sextic after elimination \\
MOSFET & bilateral square law (strong inversion); common-slope exponential law (weak inversion) & strong inversion: piecewise quadratic; weak inversion: exact \(W_0\) & strong inversion: piecewise linear or quadratic with a fixed IF common mode, Ferrari quartic with a floating IF; weak inversion: inverse of \(\sinh x+\rho x\) (fixed), \(r\)-Lambert (floating) \\
BJT/HBT & matched \(I_C=I_Se^{V_{BE}/U_E}\) & \(\tanh\) half-Gilbert; \(r\)-Lambert with emitter degeneration & \(\tanh\!\times\!\tanh\) Gilbert; nested \(r\)-Lambert with emitter degeneration \\
\bottomrule
\end{tabularx}
\end{table*}

The instantaneous solution \(v(t)\) is not yet an IF signal, since it still
contains all of the mixing products.  With \(r=\widehat r\cos\omega_Rt\) and
\(\ell=\widehat\ell\cos\omega_Lt\), we define the wanted component as
\begin{equation}
 v_{\rm IF}(t)=\mathcal P_{|\omega_R-\omega_L|}\{v(t)\},
 \label{eq:if-projection-paper}
\end{equation}
where \(\mathcal P_\omega\) denotes Fourier projection at \(\omega\) followed
by the IF transfer of the output network.  If the RF is weak enough that
\(v\simeq v_0(\ell)+r g(\ell)\), let
\begin{equation}
 G_1=\frac4\pi\int_0^{\pi/2}
 g(\widehat\ell\cos\theta)\cos\theta\,\dif\theta
 \label{eq:switching-coefficient-paper}
\end{equation}
be the first LO cosine coefficient of \(g\).  Each first-order sum and
difference sideband then has the peak value \(\widehat r|G_1|/2\) at the
core, with \(r\) defined per half-winding as above.

A matched double-balanced core is odd in \(r\) and odd in \(\ell\)
separately.  This removes the direct RF and LO terms and leaves only products
that are odd in both.  A single-balanced core lacks the second symmetry, so
its LO feedthrough is in general not canceled, although particular symmetric
passive device laws can suppress some feedthrough terms.
 \section{Semiconductor and Thermionic Diode Mixers}
\label{sec:diode-families}

\subsection{Shockley single-balanced pair: Lambert \texorpdfstring{\(W\)}{W} under one-diode dominance}

Let the two secondary terminals be at \(v_1=\ell+r\) and \(v_2=\ell-r\).
With matched Shockley diodes oriented from \(v_1\) to the output and from the
output to \(v_2\), KCL at the output is
\begin{equation}
 \frac{v}{R_L}+I_s e^{(v-v_2)/V_D}-I_se^{(v_1-v)/V_D}=0.
 \label{eq:shockley-single-dimensional-paper}
\end{equation}
The two ``\(-1\)'' terms of the diode law cancel because the diodes are
matched.  With \(x=v/V_D\), \(a=r/V_D\), \(b=\ell/V_D\), and
\(\xi=V_D/(I_sR_L)\), the full equation becomes
\begin{equation}
 e^{a+x-b}+\xi x-e^{a+b-x}=0.
 \label{eq:shockley-single-paper}
\end{equation}
Its derivative with respect to \(x\) is \(\xi\) plus both exponentials, so
there is exactly one real root.  The equation is not, however, of Lambert
form, because multiplying it by \(e^x\) leaves both an \(e^{2x}\) term and an
\(xe^x\) term.

The Lambert function appears once one diode dominates.  For \(\ell>0\) the
last exponential in \eqref{eq:shockley-single-paper} dominates and the
equation reduces to \(\xi x=e^{a+b-x}\); for \(\ell<0\) the first exponential
dominates instead.  Both cases are covered by
\begin{equation}
 v_W=sV_DW_0\!\left[
 \frac{I_sR_L}{V_D}
 \exp\!\left(\frac{|\ell|+r}{V_D}\right)\right],\quad
 s=\sgn\ell.
 \label{eq:shockley-W-paper}
\end{equation}
Because the argument is positive, the lower branch \(W_{-1}\) is not real and
there is no branch to choose.  The ratio of the neglected to the retained
diode current is \(\exp[-2|\ell-v|/V_D]\), which gives an a posteriori check:
the neglected current is at most a fraction \(\epsilon\) of the retained one
wherever
\begin{equation}
 |\ell-v|\ge \frac{V_D}{2}\ln\frac1\epsilon.
 \label{eq:shockley-W-validity-paper}
\end{equation}
Near the LO zero crossings both diodes carry comparable currents, and the
piecewise solution \eqref{eq:shockley-W-paper} jumps from one branch to the
other.  There we solve the full equation \eqref{eq:shockley-single-paper}
instead, which is easy because it is monotone.

For weak RF, let \(z_0=(I_sR_L/V_D)e^{|\ell|/V_D}\) and \(w_0=W_0(z_0)\).
Then
\begin{equation}
 v_W\simeq sV_Dw_0+s r\frac{w_0}{1+w_0},\qquad
 \frac{\partial\ln|v_W|}{\partial\ln(I_sR_L)}=\frac1{1+w_0}.
 \label{eq:shockley-small-paper}
\end{equation}
The coefficient of \(r\) is the small-signal transfer \(g(\ell)\) used in
\eqref{eq:switching-coefficient-paper}.  The second expression measures
compression.  The output grows almost in proportion to \(I_sR_L\) when
\(w_0\ll1\) and becomes nearly insensitive to it when \(w_0\gg1\), where the
RF transfer \(w_0/(1+w_0)\) approaches the ideal-switch value of unity.  This
is what allows the load and drive to be sized directly.

\subsection{Shockley double-balanced ring}

For the loaded ring with ideal voltage ratios we use the normalized variables
\(u=n_Lv_{\rm LO}/(2V_D)\), \(r=n_Rv_{\rm RF}/(2V_D)\), and
\(y=v_{\rm IF}/V_D\); note that \(r\) is normalized by \(V_D\) in this
subsection only.  Summing the four exponential branch currents gives the
single equation
\begin{equation}
 \frac{V_D}{4I_sR_L}y+\cosh u\cosh r\sinh y
 -\sinh u\sinh r\cosh y=0.
 \label{eq:shockley-double-paper}
\end{equation}
Since \(\cosh u\cosh r>|\sinh u\sinh r|\), its derivative with respect to
\(y\) is positive.  In the open-circuit limit the linear load term vanishes,
and
\begin{equation}
 y=\artanh(\tanh u\,\tanh r).
 \label{eq:shockley-double-open-paper}
\end{equation}
With a finite load, the equation contains a load-current term proportional
to \(y\) together with diode-current terms in both \(e^y\) and \(e^{-y}\).
In general it therefore cannot be rearranged into the form \(z e^z=c\) that
the ordinary Lambert \(W\) function solves.  The ring output changes sign
when either the LO or the RF voltage is reversed, so a perfectly matched ring
cancels direct LO and RF feedthrough.  With sinusoidal inputs, the remaining
output components are mixing products of odd harmonics of both signals.

For weak RF, the output is, to first order,
\begin{equation}
 y\simeq r\,
 \frac{\sinh u}{\cosh u+V_D/(4I_sR_L)},
 \label{eq:shockley-double-weak-paper}
\end{equation}
and the first LO coefficient of the factor multiplying \(r\) enters
\eqref{eq:switching-coefficient-paper}.

\subsection{Child--Langmuir single-balanced pair}

In its space-charge-limited region, a memoryless thermionic diode obeys
\begin{equation}
 \Phi_T(z)=\mathcal P\pospart{z-V_c}^{3/2},
 \label{eq:child-law-paper}
\end{equation}
where \(\mathcal P\) is the perveance and \(V_c\) is an effective knee
voltage.  Substituting this law into the single-pair circuit gives
\begin{equation}
 \frac{v}{R_L}+\mathcal P\pospart{v-\ell+r-V_c}^{3/2}
 -\mathcal P\pospart{\ell+r-V_c-v}^{3/2}=0.
 \label{eq:thermionic-single-paper}
\end{equation}
Its derivative is the load conductance plus two nonnegative incremental
conductances, each proportional to a square root, so the root is again
unique.

When only one tube conducts, let \(B=|\ell|+r-V_c>0\), \(v=sV\), and
\(\gamma=R_L\mathcal P\).  The conducting path then satisfies
\(V/R_L=\mathcal P(B-V)^{3/2}\), and the substitution
\(w=\gamma\sqrt{B-V}\) turns this into
\begin{equation}
 w^3+w^2=\eta,\qquad
 \eta=\gamma^2B,\qquad
 v=sB\frac{w}{1+w}.
 \label{eq:thermionic-cardano-paper}
\end{equation}
Since \(\eta\ge0\), this cubic has exactly one nonnegative root, which
Cardano's formula gives in closed form.  Appendix~\ref{app:stable-cardano}
lists real trigonometric and hyperbolic forms that avoid complex arithmetic.
The reduction holds as long as the second tube stays off, that is,
\(V\le |\ell|-r+V_c\); otherwise the full equation
\eqref{eq:thermionic-single-paper} is used.  Within the one-tube region, the
sensitivity of the output magnitude \(V\) to the effective drive \(B\) is
\begin{equation}
 \frac{\dif V}{\dif B}=\frac{3w}{2+3w}.
 \label{eq:thermionic-increment-paper}
\end{equation}

\subsection{Child--Langmuir double-balanced ring}

Let the ring-node voltages be \(v_5=\ell\), \(v_3=-\ell\), \(v_6=v+r\), and
\(v_4=v-r\), and let \(i_{jk}=\Phi_T(v_j-v_k)\) be the current from node
\(j\) to node \(k\).  KCL for the loaded ring is
\begin{equation}
 \begin{aligned}
 \frac v{R_L}={}&\Phi_T(v_3-v_6)-\Phi_T(v_6-v_5)\\
 &+\Phi_T(v_5-v_4)-\Phi_T(v_4-v_3).
 \end{aligned}
 \label{eq:thermionic-double-paper}
\end{equation}
Its derivative is the load conductance plus the four incremental
conductances of the conducting diodes, and the equation is odd in both \(r\)
and \(\ell\).

In the usual operating region two paths conduct.  With \(A=|\ell|-V_c\),
\(z=\sgn(\ell)v\), and \(c=(\mathcal PR_L)^{-1}\), equation
\eqref{eq:thermionic-double-paper} reduces to
\begin{equation}
 cz=(A+r-z)^{3/2}-(A-r+z)^{3/2}.
 \label{eq:thermionic-radical-paper}
\end{equation}
This reduction requires \(|r-z|<A\) and \(|r+z|\le A+2V_c\).  Both
conditions are checked on the computed root, and the reduction is used only
if they hold.  Squaring twice gives the structured sextic
\begin{equation}
 \begin{split}
 \{&(A+r-z)^3+(A-r+z)^3-c^2z^2\}^2\\
 &-4(A+r-z)^3(A-r+z)^3=0.
 \end{split}
 \label{eq:thermionic-sextic-paper}
\end{equation}
Squaring introduces extraneous roots, so for computation the monotone
radical form \eqref{eq:thermionic-radical-paper} is preferable, and the
sextic serves mainly to classify the problem algebraically.  As
\(R_L\to\infty\), the radical equation reduces to the ideal commutator
\(v=r\sgn\ell\).

Linearizing at \(r=v=0\) gives the small-signal transfer
\begin{equation}
 g_T(\ell)=\sgn\ell\,
 \frac{3\mathcal PR_L\sqrt{\pospart{|\ell|-V_c}}}
 {1+3\mathcal PR_L\sqrt{\pospart{|\ell|-V_c}}}.
 \label{eq:thermionic-gain-paper}
\end{equation}
For \(V_c=0\), let \(\zeta=3\mathcal PR_L\sqrt{\widehat\ell}\).  The
switching coefficient is then
\begin{equation}
 G_1=\frac4\pi\int_0^{\pi/2}
 \frac{\zeta\sqrt{\cos\theta}}
 {1+\zeta\sqrt{\cos\theta}}\cos\theta\,\dif\theta.
 \label{eq:thermionic-G1-paper}
\end{equation}
For \(\zeta\ll1\), Euler's beta integral gives
\begin{equation}
 G_1\simeq\frac{2\zeta}{\sqrt\pi}
 \frac{\Gamma(5/4)}{\Gamma(7/4)},
 \label{eq:thermionic-gamma-paper}
\end{equation}
whereas \(G_1\to4/\pi\) in the hard-commutating limit \(\zeta\to\infty\).
The Gamma and Beta functions thus enter only through the spectral
projection, while the Cardano cubic and the sextic come from inverting the
instantaneous KCL equation.
 \section{Depletion-FET and Vacuum-Triode Crossbars}
\label{sec:controlled-crossbars}

For three-terminal devices we use the three-balun, four-device connection of
the passive-FET mixer in \cite{MiniCircuits2015}.  We label the RF endpoints
\(a,b\), the IF endpoints \(p,n\), and the two complementary control nets
\(g_\pm\).  Writing each branch as (channel endpoint, channel endpoint,
control net), the four branches are \((a,p,g_+)\), \((a,n,g_-)\),
\((b,p,g_-)\), and \((b,n,g_+)\).  The IF center tap is held at a fixed
reference voltage, so the two IF endpoints swing equally and oppositely about
it.  With \(r\) and \(\ell\) the RF and LO half-voltages, the terminal
voltages are
\begin{equation}
 a=V_B+r,\quad b=V_B-r,\quad p=v,\quad n=-v,
 \quad g_\pm=V_G\pm\ell,
 \label{eq:crossbar-coordinates-paper}
\end{equation}
and, for a total reflected differential load \(R_\Delta\),
\begin{equation}
 \frac{4v}{R_\Delta}=I_p-I_n,\qquad
 v_{{\rm IF},o}=2n_Iv.
 \label{eq:crossbar-kcl-paper}
\end{equation}
The factor of four arises because the load sees the full differential
voltage \(2v\), while the current it draws, \(2v/R_\Delta\), is only half of
\(I_p-I_n\); the common-mode part of the injected currents returns through
the center tap.  We keep the RF center-tap bias \(V_B\) explicit.  It shifts
the channel cutoff points, so it cannot simply be set to zero.

\subsection{Matched depletion JFET/MESFET}

With \(P_J=-V_P>0\) and \(\beta_J=I_{\rm DSS}/P_J^2\), we use the
source/drain-symmetric endpoint law
\begin{equation}
 J_J(x,y;g)=\beta_J\!\left[
 \pospart{P_J+g-y}^2-\pospart{P_J+g-x}^2\right].
 \label{eq:jfet-law-paper}
\end{equation}
This is an idealized depletion-FET model.  It ignores channel-length
modulation and parasitic elements, and it is not meant to describe every
MESFET or HEMT.  For a JFET, the gate must also stay at or below both channel
terminals, \(g\le\min(x,y)\), so that both gate--channel junctions remain
reverse biased.

\begin{figure}[!t]
 \centering
 \includegraphics[width=\CircuitWidth]{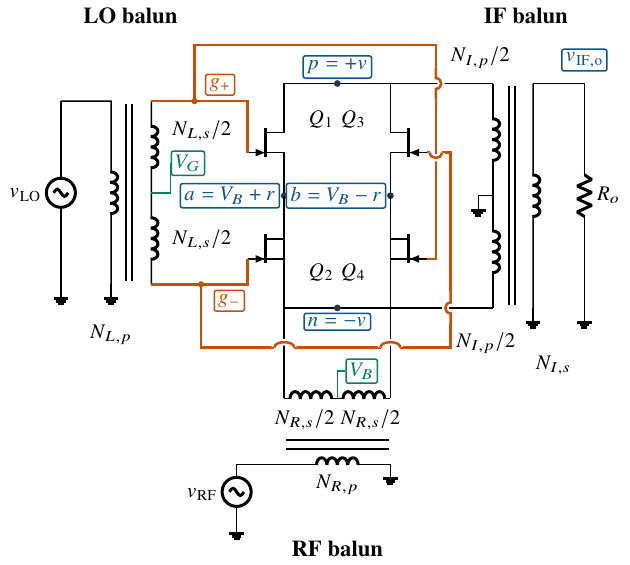}
 \caption{Parametric depletion-JFET/MESFET crossbar.  The analytical
 model uses the symbolic parameters \(I_{\rm DSS}\), \(V_P\), and
 \(R_\Delta\).}
 \label{fig:jfet-crossbar-paper}
\end{figure}

\Cref{fig:jfet-crossbar-paper} shows the complete transformer-coupled
circuit.

We define
\begin{equation}
 \begin{aligned}
 u_\pm&=P_J+V_G\pm\ell,\qquad
 \rho_J=(\beta_JR_\Delta)^{-1},\\
 h(u,x)&=\pospart{u+x}^2-\pospart{u-x}^2.
 \end{aligned}
 \label{eq:jfet-definitions-paper}
\end{equation}
Substituting the four endpoint currents into \eqref{eq:crossbar-kcl-paper}
gives the exact double-balanced equation
\begin{equation}
 \begin{aligned}
 h(u_+,v)+h(u_-,v)+4\rho_Jv\qquad\\[-1mm]
 {}=h(u_+-V_B,r)-h(u_--V_B,r).
 \end{aligned}
 \label{eq:jfet-double-paper}
\end{equation}
The derivative of its left side is
\begin{equation}
 2\!\sum_{\sigma=\pm}
 [\pospart{u_\sigma+v}+\pospart{u_\sigma-v}]+4\rho_J>0,
\end{equation}
so the root is unique.  Within a fixed conduction state, that is, for a fixed
set of active positive-part terms, \(h\) takes one of three forms.  For
\(z\ge0\) (\(h\) is odd in its second argument),
\begin{equation}
 h(u,z)=
 \begin{cases}
 0,&u\le-z,\\
 (u+z)^2,&-z<u<z,\\
 4uz,&u\ge z,
 \end{cases}
 \label{eq:jfet-branches-paper}
\end{equation}
so each candidate root solves a linear or quadratic equation.  Two closed-form
cases are particularly useful:
\begin{align}
 v&=\frac{\ell r}{U+\rho_J/2},\qquad U=P_J+V_G,
 \label{eq:jfet-all-on-paper}\\
 v&=s\frac{u_H-V_B}{u_H+\rho_J}r,\quad
 u_H=U+|\ell|,\quad s=\sgn\ell.
 \label{eq:jfet-one-diagonal-paper}
\end{align}
The first holds when all four channels conduct and the second when only one
diagonal conducts.  In both cases the positive-part inequalities must be
checked once the root has been evaluated.

The natural single-balanced counterpart uses two devices and a load \(R_S\).
With \(\rho_{J,S}=(\beta_JR_S)^{-1}\), KCL gives
\begin{equation}
 \begin{aligned}
 \rho_{J,S}v={}&\pospart{u_+-v}^2+\pospart{u_--v}^2\\
 &-\pospart{u_+-V_B-r}^2-\pospart{u_--V_B+r}^2.
 \end{aligned}
 \label{eq:jfet-single-paper}
\end{equation}
Let \(\mathcal A\) be the set of active output-side terms, with
\(m=|\mathcal A|\) elements, and let \(S=\sum_\mathcal A u_\sigma\),
\(C=\sum_\mathcal A u_\sigma^2\), and \(D\) be the sum of the two input-side
positive-part squares in \eqref{eq:jfet-single-paper}.  For \(m\ge1\), the
root that is continuous within this conduction state is
\begin{equation}
 v=\frac{2S+\rho_{J,S}-
 \sqrt{(2S+\rho_{J,S})^2-4m(C-D)}}{2m},
 \label{eq:jfet-single-root-paper}
\end{equation}
and for \(m=0\) it is simply \(v=-D/\rho_{J,S}\).  As before, each candidate
is accepted only if it satisfies the inequalities that define its conduction
state.  The depletion FET therefore needs no Lambert-type function at all,
since its solution is piecewise quadratic and involves only ordinary square
roots.  Arctangent or hyperbolic functions can appear, but only later, when
the piecewise waveform is Fourier projected.

For weak RF, implicit differentiation of \eqref{eq:jfet-double-paper} gives
the double-balanced transfer coefficient directly:
\begin{equation}
 g_{J,D}(\ell)=
 \frac{\pospart{u_+-V_B}-\pospart{u_--V_B}}
 {\pospart{u_+}+\pospart{u_-}+\rho_J}.
 \label{eq:jfet-double-gain-paper}
\end{equation}
For the single-balanced circuit, with \(v_0(\ell)\) the zero-RF root of
\eqref{eq:jfet-single-paper}, the corresponding coefficient is
\begin{equation}
 g_{J,S}(\ell)=
 \frac{\pospart{u_+-V_B}-\pospart{u_--V_B}}
 {\pospart{u_+-v_0}+\pospart{u_--v_0}+\rho_{J,S}/2}.
 \label{eq:jfet-single-gain-paper}
\end{equation}
Let \(G_{1,\chi}\) be the first LO coefficient of \(g_{J,\chi}\), defined as
in \eqref{eq:switching-coefficient-paper}.  Since \(r=n_Rv_{\rm RF}/2\), the
external first-order sideband for \(\chi\in\{S,D\}\) has the peak value
\begin{equation}
 \widehat V_{{\rm IF},o,\chi}=n_I\widehat r|G_{1,\chi}|
 =\frac{n_In_R}{2}\widehat V_{\rm RF}|G_{1,\chi}|,
 \quad \chi\in\{S,D\}.
 \label{eq:jfet-final-if-paper}
\end{equation}

\subsection{Vacuum-triode crossbar with complete biasing}

\begin{figure}[!t]
 \centering
 \includegraphics[width=\CircuitWidth]{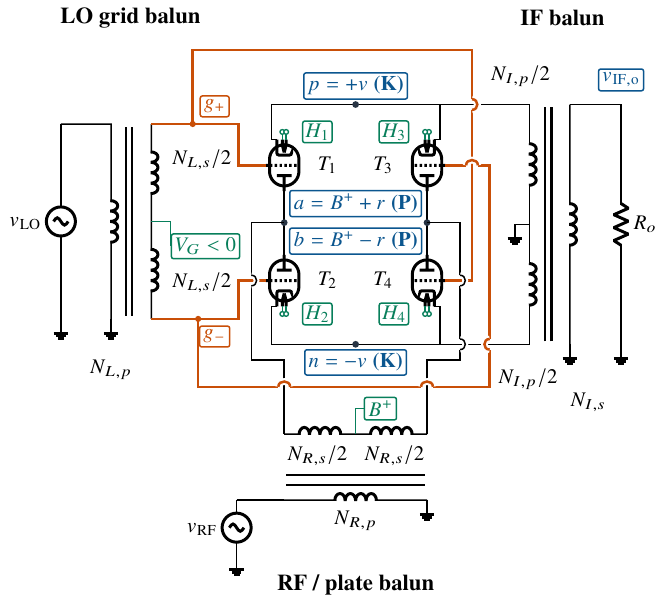}
 \caption{Parametric four-triode crossbar, including the plate and cathode
 connections, the plate supply \(B^+\), the negative grid bias, the IF
 return, and the isolated heater supplies.  The analytical model uses the
 symbolic parameters \(\mathcal P_3\), \(\mu\), \(V_c\), and \(R_\Delta\).}
 \label{fig:triode-crossbar-paper}
\end{figure}

Unlike a FET channel, a triode conducts in one direction only, so it cannot
simply replace a FET in the crossbar.  To keep the same connection graph, the
plates form the RF endpoints and are supplied through the RF center tap
(\(B^+=V_B\)), the cathodes form the IF endpoints with a common current
return, the grids are driven in antiphase about a negative bias, and the
heaters are isolated.  The circuit in
\cref{fig:triode-crossbar-paper} is therefore a biased triode mixer with the
IF taken from the cathodes, not a device-for-device replacement of the FETs
in a passive mixer.

We assume that the grids stay negative, so that no grid current flows, and
use the three-halves-power law
\begin{equation}
 \begin{aligned}
 I_3(P,K,G)&=\Phi_3(E),\qquad
 \Phi_3(E)=\mathcal P_3\pospart{E}^{3/2},\\
 E&=V_{GK}+\frac{V_{PK}}\mu-V_c.
 \end{aligned}
 \label{eq:triode-law-paper}
\end{equation}
With \(A=V_G+V_B/\mu-V_c\) and \(\kappa=1+1/\mu\), the controls of the four
branches are
\begin{align}
 E_{pa}&=A+\ell+r/\mu-\kappa v,&
 E_{na}&=A-\ell+r/\mu+\kappa v,\notag\\
 E_{pb}&=A-\ell-r/\mu-\kappa v,&
 E_{nb}&=A+\ell-r/\mu+\kappa v,
 \label{eq:triode-controls-paper}
\end{align}
where the first subscript names the cathode and the second the plate.
Equation \eqref{eq:crossbar-kcl-paper} then becomes
\begin{equation}
 \frac{4v}{R_\Delta}=\Phi_3(E_{pa})+\Phi_3(E_{pb})
 -\Phi_3(E_{na})-\Phi_3(E_{nb}).
 \label{eq:triode-double-paper}
\end{equation}
With all terms moved to the left, the derivative with respect to \(v\) is
\(4/R_\Delta+\kappa\sum\Phi_3'(E)>0\).  The root is therefore unique,
including across every cutoff transition, and for matched tubes it is odd in
both drives.

At \(r=v=0\), let \(G_\pm=(3\mathcal P_3/2)\pospart{A\pm\ell}^{1/2}\) be the
incremental conductances.  The weak-RF double-balanced coefficient is then
\begin{equation}
 g_{3,D}(\ell)=
 \frac{(G_+-G_-)/\mu}{2/R_\Delta+\kappa(G_++G_-)},
 \label{eq:triode-double-gain-paper}
\end{equation}
and its first LO coefficient gives the external IF through
\eqref{eq:jfet-final-if-paper}.  The single-triode coefficient has the same
form, with \(2/R_\Delta\) replaced by \(1/R_S\) and the incremental
conductances evaluated at the loaded zero-RF operating point.

When the LO is strong enough that only one diagonal conducts, let
\(s=\sgn\ell\), \(z=sv\), \(H=A+|\ell|\), \(\delta=r/\mu-\kappa z\), and
\(D=4z/(\mathcal P_3R_\Delta)\).  The equation becomes
\begin{equation}
 D=(H+\delta)^{3/2}-(H-\delta)^{3/2}.
 \label{eq:triode-radical-paper}
\end{equation}
This reduction requires \(H+\delta>0\), \(H-\delta>0\), and both controls on
the opposite diagonal to be at or below cutoff.  As before, these conditions
are checked on the computed root.  Eliminating the radicals gives
\begin{equation}
 [(H+\delta)^3+(H-\delta)^3-D^2]^2
 -4(H^2-\delta^2)^3=0.
 \label{eq:triode-sextic-paper}
\end{equation}
Because \(D\) and \(\delta\) are affine in \(z\), this is in general a sextic
in \(z\).  In the open-circuit limit the solution is elementary,
\(v=r\sgn(\ell)/(\mu+1)\).

In the two-triode single-balanced circuit, both cathodes feed one loaded
node, so
\begin{equation}
 \frac v{R_S}=\Phi_3\!\left(A+\ell+\frac r\mu-\kappa v\right)
 +\Phi_3\!\left(A-\ell-\frac r\mu-\kappa v\right).
 \label{eq:triode-single-paper}
\end{equation}
When only one tube conducts, let \(B_s=A+|\ell|+sr/\mu\),
\(\gamma=\kappa\mathcal P_3R_S\), and \(w=\gamma\sqrt{B_s-\kappa v}\).  The
same Cardano reduction as for the thermionic diode then follows:
\begin{equation}
 w^3+w^2=\gamma^2B_s,\qquad
 v=\frac{B_s}{\kappa}\frac{w}{1+w}.
 \label{eq:triode-cardano-paper}
\end{equation}
The three-halves exponent thus fixes the algebra.  A single conducting tube
leads to a Cardano cubic, and the loaded double core leads to a structured
sextic.  In practice, the plate must also stay above the cathode and the grid
negative, emission and grid-current limits must be respected, and heater and
interelectrode-capacitance effects need separate models.
 \section{MOSFET Mixers with Different IF Load and Bias Connections}
\label{sec:mosfet-paper}

A passive MOS crossbar needs a channel that is symmetric between source and
drain, and a body connection that keeps both source/drain-to-body junctions
reverse biased.  A discrete three-terminal NMOS, whose body is tied
internally to one channel terminal, is therefore not by itself the bilateral
element assumed here.  As before, \(r\), \(\ell\), and \(v\) are the RF, LO
(gate), and IF half-differential voltages, and \(g_\pm=V_G\pm\ell\).  The
single-balanced core connects \(\pm r\) to one loaded node, and the
double-balanced core uses the four-device crossbar of
\cref{fig:generic-topologies-paper}.

\subsection{Strong-inversion bilateral square law}

We describe each channel by the symmetric endpoint potential \(\Xi_g\) and
the signed channel current \(J_\square\),
\begin{equation}
 \Xi_g(x)=\frac\beta2\pospart{g-V_{T0}-x}^2,\qquad
 J_\square(a,b;g)=\Xi_g(b)-\Xi_g(a).
 \label{eq:mos-square-law-paper}
\end{equation}
This is a simplified long-channel form of the Shichman--Hodges model
\cite{ShichmanHodges1968}.  Body effect, channel-length modulation, mobility
reduction, velocity saturation, and terminal charge are neglected.

The full single-balanced equation is
\begin{equation}
 \frac v{R_L}=J_\square(r,v;g_+)+J_\square(-r,v;g_-).
 \label{eq:mos-single-square-paper}
\end{equation}
With \(u_\pm=g_\pm-V_{T0}\), the derivative of the left side minus the right
side is \(R_L^{-1}+\beta\pospart{u_+-v}+\beta\pospart{u_--v}>0\).  For every
fixed pattern of active positive-part terms, the equation is at most
quadratic.  If all channels remain in the triode region, then with
\(U=V_G-V_{T0}\),
\begin{equation}
 \frac v{R_L}=\beta(v^2-2Uv+2\ell r-r^2),
\end{equation}
and the root that is continuous from \(v=0\) is
\begin{equation}
 v=\frac{A-\sqrt{A^2-8\ell r+4r^2}}2,\qquad
 A=2U+(\beta R_L)^{-1}.
 \label{eq:mos-single-square-root-paper}
\end{equation}
The global solution is assembled from the roots for zero, one, and two
active output-side terms, discarding any candidate that violates the
conduction state it was derived for.  Appendix~\ref{app:mos-global} gives a
numerically stable explicit form.  In strong inversion, then, the
single-balanced MOS mixer needs nothing beyond ordinary square roots.

When the IF common-mode voltage is held at a fixed reference, so that
\(p=v\) and \(n=-v\), differential KCL gives, with the function \(h\) of
\eqref{eq:jfet-definitions-paper},
\begin{equation}
 h(u_+,v)+h(u_-,v)+\frac{8v}{\beta R_L}
 =h(u_+,r)-h(u_-,r).
 \label{eq:mos-double-clamped-paper}
\end{equation}
Because \(\partial h/\partial x=2\pospart{u+x}+2\pospart{u-x}\) is
nonnegative, the root is unique, and each fixed conduction state again gives a
linear or quadratic equation.  If all four channels remain in the triode
region,
\begin{equation}
 v=\frac{\ell r}{U+1/(\beta R_L)}.
 \label{eq:mos-double-all-on-paper}
\end{equation}

The published three-balun FET circuit \cite{MiniCircuits2015} fixes the LO
and IF center-tap voltages but leaves the RF center-tap bias unspecified.  If
the RF and IF common-mode voltages differ by \(\Delta_B\), the equation
becomes
\begin{equation}
 \begin{aligned}
 h(\bar u_+,v)+h(\bar u_-,v)+\frac{8v}{\beta R_L}\qquad\\[-1mm]
 {}=h(\bar u_+-\Delta_B,r)-h(\bar u_--\Delta_B,r),
 \end{aligned}
 \label{eq:mos-an003-paper}
\end{equation}
where \(\bar u_\pm=C_G-C_I-V_{T0}\pm\ell\).  Setting \(\Delta_B=0\) without
knowing the actual center-tap voltage can change the predicted waveform
substantially.  In the test case of \cref{sec:validation-paper}, the error is
32.95\% NRMSE.

Another possible IF connection uses a resistor \(R_h\) from each IF endpoint
to the same fixed reference.  Let \(P\) and \(N\) be the endpoint voltages
relative to that reference.  The resistors do not force \(P+N=0\), and each
endpoint obeys its own node equation:
\begin{equation}
 \begin{aligned}
 \frac{P}{R_h}
 &=J_\square(r,P;g_+)+J_\square(-r,P;g_-),\\
 \frac{N}{R_h}
 &=J_\square(-r,N;g_+)+J_\square(r,N;g_-).
 \end{aligned}
 \label{eq:mos-half-load-paper}
\end{equation}
Each equation has a unique, piecewise-quadratic root, and the differential
and common-mode voltages follow as \(2v=P-N\) and \(c=(P+N)/2\).  The
resistors give the common mode a return path to the reference without fixing
its value.  This circuit therefore differs both from a center-tapped winding
held at a fixed voltage and from an isolated floating winding.

For small RF, the fixed-common-mode circuit and the floating circuit
discussed next have the same first-order transfer coefficient,
\begin{equation}
 g_\square(\ell)=
 \frac{\pospart{U+\ell}-\pospart{U-\ell}}
 {\pospart{U+\ell}+\pospart{U-\ell}+2/(\beta R_L)},
 \label{eq:mos-square-gain-paper}
\end{equation}
provided the floating circuit stays on the conducting solution for which
\(c\to0\) as \(r\to0\).  The coefficient is odd in \(\ell\), and its first LO
coefficient \eqref{eq:switching-coefficient-paper} gives the first-order sum
and difference tones.

For an isolated floating IF winding, we set \(p=c+v\) and \(n=c-v\).  The two
node equations are
\begin{equation}
 \begin{aligned}
 \frac{2v}{R_L}&=J_\square(r,c+v;g_+)+J_\square(-r,c+v;g_-),\\
 -\frac{2v}{R_L}&=J_\square(-r,c-v;g_+)+J_\square(r,c-v;g_-).
 \end{aligned}
 \label{eq:mos-floating-pair-paper}
\end{equation}
When all channels remain in the triode region, let \(q=U-c\) and
\(a=(\beta R_L)^{-1}\).  Adding and subtracting the two equations gives
\begin{equation}
 q^2+v^2=U^2+r^2,\qquad v(a+q)=\ell r.
 \label{eq:mos-floating-reduced-paper}
\end{equation}
Eliminating \(q\) yields the quartic
\begin{equation}
 v^4-(U^2+r^2-a^2)v^2-2a\ell r\,v+\ell^2r^2=0,
 \label{eq:mos-ferrari-paper}
\end{equation}
which can be solved in closed form by Ferrari's method.  The physical root is
the one that satisfies both equations in \eqref{eq:mos-floating-pair-paper}
and all four triode inequalities.  The other conduction states also lead to
polynomials of degree at most four.  For numerical work it is often more
robust to solve \eqref{eq:mos-floating-pair-paper} directly, while the
quartic gives the same root in closed form.

\subsection{Weak inversion with ordinary and generalized Lambert functions}

In weak inversion we use the source/drain-symmetric exponential law
\begin{equation}
 J_w(a,b;g)=I_0e^{(g-V_{T0})/(nU_T)}
 \left(e^{-b/U_T}-e^{-a/U_T}\right),
 \label{eq:mos-weak-law-paper}
\end{equation}
in which every channel endpoint enters with the same slope \(1/U_T\).  Let
\(A_\pm=I_0e^{(g_\pm-V_{T0})/(nU_T)}\), \(\rho=r/U_T\), and \(x=v/U_T\).  KCL
at the loaded node of the single-balanced circuit becomes
\begin{equation}
 \begin{aligned}
 x&=C e^{-x}-D,\qquad C=\frac{R_L(A_++A_-)}{U_T},\\
 D&=\frac{R_L(A_+e^{-\rho}+A_-e^\rho)}{U_T},
 \end{aligned}
\end{equation}
and its exact solution, which retains both channel currents, is
\begin{equation}
 v=U_T[W_0(Ce^D)-D].
 \label{eq:mos-weak-W-paper}
\end{equation}
Unlike the Shockley result \eqref{eq:shockley-W-paper}, this \(W_0\) is not an
approximation, because no current has been neglected.  It is exact because
both output-dependent currents depend on \(v\) through the same factor
\(e^{-v/U_T}\) and therefore combine into a single exponential.  For the
devices actually to be in weak inversion, the pinch-off potential
\(V_{P,j}\) must satisfy \(\exp[(V_{P,j}-z)/(2U_T)]\ll1\) at every channel
endpoint \(z\).  Outside this region, \eqref{eq:mos-weak-W-paper} is still
the exact solution of the exponential law, but that law no longer describes a
real MOSFET accurately.

If the IF common mode of the double-balanced bridge is held at a fixed
reference, the differential KCL equation is
\begin{equation}
 (A_++A_-)\sinh(v/U_T)+\frac{2v}{R_L}
 =(A_+-A_-)\sinh(r/U_T),
\end{equation}
or, after normalization,
\begin{equation}
 \begin{aligned}
 \sinh x+\rho_cx&=q_c,\\
 \rho_c&=\frac{2U_T}{R_L(A_++A_-)},\\
 q_c&=\frac{A_+-A_-}{A_++A_-}\sinh\rho.
 \end{aligned}
 \label{eq:mos-weak-clamped-paper}
\end{equation}
The left side has derivative \(\cosh x+\rho_c>0\), so the solution is the
unique inverse of the monotone map \(x\mapsto\sinh x+\rho_cx\).  For a finite
load, this inverse is neither the ordinary \(W\) function nor the standard
\(r\)-Lambert function.  In the open-circuit limit \(\rho_c=0\), it reduces
to \(x=\asinh q_c\).

For the floating bridge we write
\begin{equation}
 p=U_T(\xi+y),\quad n=U_T(\xi-y),\quad
 s_\pm=\frac{R_LA_\pm}{2U_T},
\end{equation}
and define \(d_+=s_+e^{-\rho}+s_-e^\rho\) and
\(d_-=s_+e^\rho+s_-e^{-\rho}\).  Eliminating the common-mode variable \(\xi\)
from the two node equations leaves the single equation
\begin{equation}
 (y+d_+)e^{2y}=d_--y.
 \label{eq:mos-weak-floating-reduced-paper}
\end{equation}
The physical solution lies in the interval \(-d_+<y<d_-\).  There the
residual \((y+d_+)e^{2y}+y-d_-\) has derivative
\(e^{2y}[1+2(y+d_+)]+1>0\), so the root is unique.  The real \(r\)-Lambert
function is defined by \(z e^z+rz=a\Longleftrightarrow z=\mathcal W_r(a)\)
\cite{MezoBaricz2017}, and the substitution \(z=2(y+d_+)\) gives
\begin{equation}
 y=\frac12\rLambert{e^{2d_+}}
 {2(d_++d_-)e^{2d_+}}-d_+.
 \label{eq:mos-weak-rLambert-paper}
\end{equation}
Letting the IF common mode float, instead of holding it at a fixed reference,
thus changes the solution from the inverse of \(\sinh x+\rho x\) to the real
\(r\)-Lambert function, even though the four channel laws are exactly the
same.

For small RF, the fixed-common-mode and floating circuits again have the same
first-order transfer coefficient,
\begin{equation}
 g_w(\ell)=\frac{\chi\sinh\lambda}{1+\chi\cosh\lambda},\qquad
 \chi=\frac{R_LA_0}{U_T},\quad \lambda=\frac\ell{nU_T}.
\end{equation}
With \(\lambda=\Lambda\cos\theta\), its first LO coefficient has the
convergent modified-Bessel representation
\begin{equation}
 G_1=4\sum_{k=1}^{\infty}(-1)^{k+1}q_B^kI_1(k\Lambda),\quad
 q_B=\frac{1-\sqrt{1-\chi^2}}\chi,
 \label{eq:mos-bessel-paper}
\end{equation}
provided \(0<\chi<1\) and \(q_Be^\Lambda<1\).  Otherwise the Fourier
integral \eqref{eq:switching-coefficient-paper}, whose integrand is bounded,
is the better choice.

\subsection{Active MOS differential pair}

The active source-coupled pair is a different circuit from the passive
crossbar.  We include it because its spectral analysis leads to complete
elliptic integrals.  For two saturated square-law devices sharing a tail
current \(I_T\), let \(z=v_d/[2\sqrt{I_T/\beta}]\).  The normalized current
difference is
\begin{equation}
 \frac{i_1-i_2}{I_T}=\begin{cases}
 2z\sqrt{1-z^2},&|z|\le1/\sqrt2,\\
 \sgn z,&|z|>1/\sqrt2.
 \end{cases}
 \label{eq:mos-active-transfer-paper}
\end{equation}
For an unclipped sinusoidal drive \(z=m\cos\theta\), that is, with
\(m\le1/\sqrt2\), the first cosine coefficient is
\begin{equation}
 S_1(m)=\frac{8}{3\pi m}
 [(2m^2-1)E(m)+(1-m^2)K(m)],
 \label{eq:mos-elliptic-paper}
\end{equation}
where \(K\) and \(E\) are the complete elliptic integrals of the first and
second kind.  Here the elliptic integrals come from the Fourier projection of
the output current, not from solving a current-balance equation as in the
passive mixers.
 \section{BJT and HBT Current-Steering Mixers}
\label{sec:bipolar-paper}

Bipolar transistors are unilateral, so instead of substituting them into a
passive ring we place them in transformer-driven active cells: a half-Gilbert
cell for single-balanced operation and a Gilbert cell for double-balanced
operation.  For matched forward-active devices,
\begin{equation}
 I_C=I_Se^{V_{BE}/U_E},
\end{equation}
and a differential pair with available collector current \(J\) divides it
according to
\begin{equation}
 I_1-I_2=J\tanh\!\left(\frac{v_d}{2U_E}\right).
 \label{eq:bjt-pair-paper}
\end{equation}
As before, \(r=n_Rv_{\rm RF}/2\) and \(\ell=n_Lv_{\rm LO}/2\) already include
the transformer scaling.  Let \(U_R\) and \(U_L\) be the effective
exponential voltage scales of the lower (RF) and upper (LO) pairs; for
matched ideal devices both equal \(U_E\).  With a total available collector
current \(I_T\), the lower RF pair delivers the branch currents
\begin{equation}
 J_{a,b}=\frac{I_T}{2}\left[1\pm\tanh(r/U_R)\right].
 \label{eq:bjt-rf-branches-paper}
\end{equation}

\subsection{Single- and double-balanced cells}

\begin{figure}[!t]
 \centering
 \includegraphics[width=\CircuitWidth]{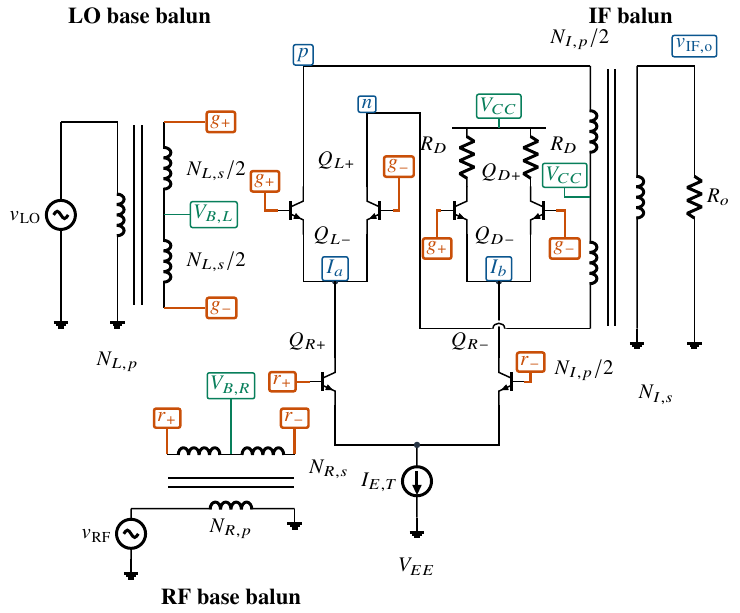}
 \caption{Transformer-coupled single-balanced half-Gilbert cell with six
 matched BJTs or HBTs.  The extra (dummy) upper pair carries the RF branch
 current that does not reach the IF output and keeps the corresponding lower
 transistor in the forward-active region.  The symbolic parameters are
 \(I_{E,T}\), \(R_D\), \(R_o\), \(n_R\), \(n_L\), and \(n_I\).}
 \label{fig:bjt-single-paper}
\end{figure}

In the single-balanced cell, only branch \(J_a\) is commutated by an LO pair.
Branch \(J_b\) is routed through a matched dummy pair, so the observed output
is
\begin{equation}
 i_{\Delta,S}=\frac{I_T}{2}
 [1+\tanh(r/U_R)]\tanh(\ell/U_L).
 \label{eq:bjt-single-paper}
\end{equation}
\Cref{fig:bjt-single-paper} shows the complete transformer-coupled circuit.

The product term in \eqref{eq:bjt-single-paper} is balanced with respect to
the RF, but the leading term \((I_T/2)\tanh(\ell/U_L)\) is direct LO
feedthrough, which the single-balanced cell cannot cancel.  The IF
transformer converts the differential current into
\begin{equation}
 v_{{\rm IF},o}=-\frac{n_IR_\Delta}{2}i_\Delta,
 \label{eq:bjt-output-map-paper}
\end{equation}
where \(R_\Delta\) is the full reflected differential resistance.

\begin{figure}[!t]
 \centering
 \includegraphics[width=\CircuitWidth]{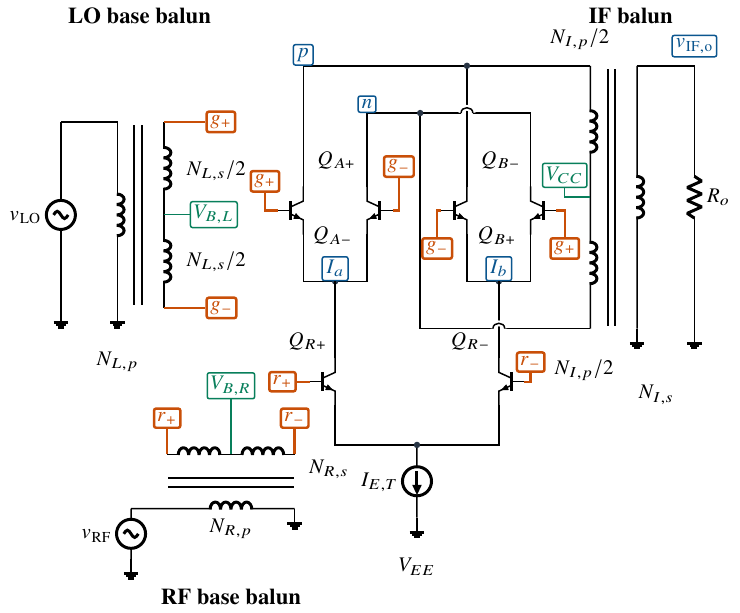}
 \caption{Transformer-coupled six-transistor double-balanced Gilbert cell
 with matched BJTs or HBTs.  The winding dots determine the sign of the
 external IF voltage.  The symbolic parameters are \(I_{E,T}\), \(R_o\),
 \(n_R\), \(n_L\), and \(n_I\).}
 \label{fig:bjt-double-paper}
\end{figure}

Commutating both lower-pair branches with cross-coupled LO pairs gives the
six-transistor double-balanced Gilbert cell, whose output is
\begin{equation}
 i_{\Delta,D}=I_T\tanh(r/U_R)\tanh(\ell/U_L),
 \label{eq:bjt-double-paper}
\end{equation}
again converted to a voltage by \eqref{eq:bjt-output-map-paper}.  This output
is odd in the RF and in the LO separately.  In an ideal matched core,
therefore, the direct RF and LO feedthrough cancel, as do all products that
involve an even harmonic of either input.  \Cref{fig:bjt-double-paper} shows
the corresponding transformer-coupled circuit.

For a sinusoidal drive with normalized peak \(a\), let \(T_1(a)\) be the
first cosine coefficient of \(\tanh(a\cos\theta)\):
\begin{equation}
 T_1(a)=\frac4\pi\int_0^{\pi/2}
 \tanh(a\cos\theta)\cos\theta\,\dif\theta.
 \label{eq:bjt-T1-paper}
\end{equation}
Provided no higher-order product falls on the same frequency, each of the
first-order difference and sum tones then has the peak value
\begin{equation}
 \begin{aligned}
 \widehat V_{{\rm IF},S}^{(1,1)}
 &=\frac{n_IR_\Delta I_T}{8}T_1(a_R)T_1(a_L),\\
 \widehat V_{{\rm IF},D}^{(1,1)}
 &=\frac{n_IR_\Delta I_T}{4}T_1(a_R)T_1(a_L),
 \end{aligned}
 \label{eq:bjt-sidebands-paper}
\end{equation}
where \(a_R=\widehat r/U_R\) and \(a_L=\widehat\ell/U_L\).  The small-drive
expansion of \(T_1\) and its large-drive limit (\(a\to\infty\)) are
\begin{equation}
 T_1(a)=a-\frac{a^3}{4}+\frac{a^5}{12}+O(a^7),\qquad
 T_1(a)\to\frac4\pi.
 \label{eq:bjt-T1-limits-paper}
\end{equation}

\subsection{Exact \texorpdfstring{\(r\)}{r}-Lambert steering with equal
emitter degeneration}

Now let each emitter of a pair have a resistance \(R_E\), let the pair carry
a total emitter current \(J_E\), and define the normalized current difference
\(m=(I_{E1}-I_{E2})/J_E\).  The differential input voltage is then exactly
\begin{equation}
 v_d=2U_E\artanh m+R_EJ_Em.
\end{equation}
With \(a=v_d/(2U_E)\) and \(b=R_EJ_E/(2U_E)\), this becomes the implicit
equation
\begin{equation}
 m=\tanh(a-bm).
 \label{eq:bjt-degenerated-implicit-paper}
\end{equation}
The substitution \(w=2b(m+1)\) turns it into
\(we^w+e^{2(a+b)}w=4be^{2(a+b)}\), which is the defining equation of the
\(r\)-Lambert function.  Hence
\begin{equation}
 \mathscr S(a,b)=
 \begin{cases}
 \tanh a,&b=0,\\[1mm]
 \displaystyle\frac{1}{2b}
 \rLambert{e^{2(a+b)}}{4be^{2(a+b)}}-1,&b>0.
 \end{cases}
 \label{eq:bjt-rLambert-paper}
\end{equation}
Because the residual \(\artanh m+bm-a\) is strictly increasing in \(m\), this
is the only physical solution.

In the degenerated mixer, the RF pair divides its current as
\(J_{a,b}=I_T(1\pm x_E)/2\), where \(x_E=\mathscr S(a_R,b_R)\) replaces
\(\tanh(r/U_R)\) in \eqref{eq:bjt-rf-branches-paper}.  Each upper pair then
divides its own branch current according to
\(s_{a,b}=\mathscr S(a_L,R_{E,L}J_{a,b}/(2U_L))\), which gives
\begin{equation}
 v_{{\rm IF},o,S}=-\frac{n_IR_\Delta}{2}J_as_a,\qquad
 v_{{\rm IF},o,D}=-\frac{n_IR_\Delta}{2}(J_as_a-J_bs_b).
 \label{eq:bjt-degenerated-mixers-paper}
\end{equation}
These expressions assume that every transistor stays in the forward-active
region and that base currents are negligible (infinite current gain).  They
remain valid for finite current gain if \(J_a\) and \(J_b\) are taken to be
the collector currents actually available to the upper pairs.
Alternatively, if finite current gain is modeled explicitly, the factors
\(\alpha=I_C/I_E\) must be applied wherever emitter currents are converted
to collector currents.  With degeneration, the current division is solved
twice in succession, first for the lower RF pair and then for each upper LO
pair, and this is what produces the nested \(r\)-Lambert functions.  They
are a consequence of the feedback through the emitter resistors, not of the
heterojunction structure of an HBT.

Under the same matched, forward-active assumptions, an HBT obeys the same
normalized exponential and \(\tanh\) relations as a BJT.  What differs from
one device to another are \(I_S\), the effective ideality factor, the current
gain, the Early effect, charge storage, and the high-frequency response
\cite{Kroemer1957,GummelPoon1970}.  \Cref{sec:validation-paper} compares the
equations with a detailed model of one commercial HBT, the NXP BFU760F, at
low frequency \cite{NXPBFU760F}.
 \begin{figure}[!t]
 \centering
 \includegraphics[width=\PlotWidth]{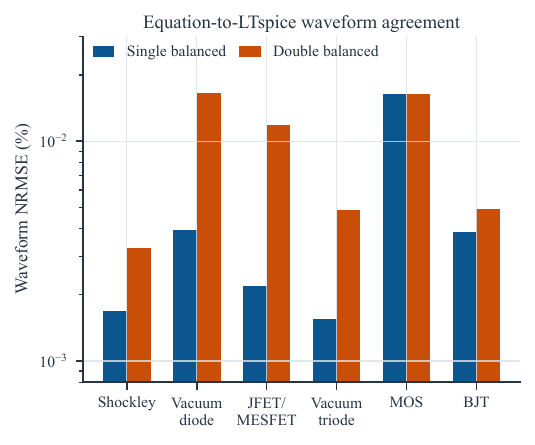}
 \caption{Waveform NRMSE between the analytical and LTspice IF outputs for
 one single-balanced and one double-balanced mixer of each device family
 (logarithmic scale).  All twelve values are below \(0.017\%\).  Results
 obtained with LTspice's built-in device elements and with the detailed
 BFU760F model are listed separately in \cref{tab:headline-validation-paper}.}
 \label{fig:validation-summary-paper}
\end{figure}

\section{LTspice Validation}
\label{sec:validation-paper}

We tested the analytical results for all six device families with 37 LTspice
netlists, each tied to a specific mixer equation or to a supplementary check.
Seven Python programs read the simulated waveforms and compared them with the
equations, both in the time domain and at the relevant FFT bins.  Where an
equation holds only under stated operating conditions, the program for that
device family also checked those conditions and used only the samples that
satisfied them.  A separate driver script ran all netlists in sequence and
confirmed that each simulation finished without error and wrote fresh output
files (Appendix~\ref{app:reproducibility-paper}).

We quantify the agreement by the normalized root-mean-square error
\begin{equation}
 {\rm NRMSE}=100\,
 \frac{\sqrt{N^{-1}\sum_k(v_{{\rm LT},k}-v_{{\rm eq},k})^2}}
 {\sqrt{N^{-1}\sum_kv_{{\rm eq},k}^2}}\ \%,
\end{equation}
where \(N\) is the number of samples and \(v_{{\rm LT},k}\) and
\(v_{{\rm eq},k}\) are the LTspice and analytical voltages at sample \(k\).
In other words, the NRMSE is the RMS difference between the two waveforms,
expressed as a percentage of the RMS value of the analytical waveform.
Because LTspice stores the transient solution at nonuniform time steps, each
result was interpolated once onto a uniform grid spanning an integer number
of RF and LO cycles.  This places the frequencies of interest exactly on FFT
bins.  No amplitude scaling or fitted gain was applied before the NRMSE was
computed.

\Cref{fig:validation-summary-paper} shows the twelve primary comparisons,
and \cref{tab:headline-validation-paper} lists the corresponding values
together with the LTspice model used in each case.

\begin{table*}[!t]
\caption{Waveform comparison for the complete single- and double-balanced
mixers.  The second column describes how each device law was implemented in
LTspice, and the last column summarizes an additional simulation or
operating-condition check.}
\label{tab:headline-validation-paper}
\centering
\footnotesize
\setlength{\tabcolsep}{4.5pt}
\begin{tabularx}{\textwidth}{@{}
 >{\raggedright\arraybackslash}p{0.16\textwidth}
 >{\raggedright\arraybackslash}p{0.30\textwidth}
 r r
 >{\raggedright\arraybackslash}X@{}}
\toprule
Device law & LTspice implementation & \shortstack{Single-balanced\\NRMSE} &
\shortstack{Double-balanced\\NRMSE} & Additional simulation or check \\
\midrule
Shockley junction & behavioral current sources implementing the full matched Shockley law & 0.001688\% & 0.003258\% & spectra of the complete two-diode pair and of the finite-load ring \\
Child--Langmuir diode & behavioral current sources implementing the \(3/2\)-power Child--Langmuir law & 0.003929\% & 0.016500\% & adds plate-to-cathode capacitance \(C_{AK}\) and near-ideal coupled windings \\
Depletion JFET/MESFET & behavioral current sources implementing the bidirectional endpoint law & 0.002198\% & 0.011830\% & LTspice Level-1 JFET: 0.00180\% (single) and 0.00781\% (double) \\
Vacuum triode & \(3/2\)-power triode law entered directly, with plate and grid bias & 0.001551\% & 0.004877\% & confirms that the grid-to-cathode and plate-to-cathode voltages stay in the assumed ranges \\
Passive MOS & bidirectional square-law channel equation entered directly & 0.016417\% & 0.016438\% & coupled-winding and Level-1 runs use the resistive IF connection of \eqref{eq:mos-half-load-paper}, which does not force the two IF endpoint voltages to be equal and opposite \\
BJT & LTspice NPN elements with near-ideal forward-active parameters & 0.003860\% & 0.004910\% & detailed BFU760F HBT model: 1.402\% (single) and 2.794\% (double) \\
\bottomrule
\end{tabularx}
\end{table*}

We also checked each special-function and polynomial solution against a
direct numerical solution of the original current-balance equation.  This is
an independent test, because the numerical solver starts from the circuit
equation rather than from the rearranged closed form.
\Cref{tab:closed-form-validation-paper} summarizes representative results.
The Shockley \(W_0\) entry is treated differently, because that result is an
approximation based on one-diode dominance.  For it we included only the
samples at which the neglected diode current was at most 1\% of the retained
current, that is, \(\epsilon=0.01\) in \eqref{eq:shockley-W-validity-paper}.

\begin{table*}[!t]
\caption{Checks of the analytical solutions and of the assumptions used in
their derivation.  The third column states the comparison made in each case.}
\label{tab:closed-form-validation-paper}
\centering
\footnotesize
\setlength{\tabcolsep}{4.2pt}
\begin{tabularx}{\textwidth}{@{}
 >{\raggedright\arraybackslash}p{0.20\textwidth}
 >{\raggedright\arraybackslash}p{0.26\textwidth}
 >{\raggedright\arraybackslash}X@{}}
\toprule
Mixer configuration & Analytical result & Comparison result \\
\midrule
Single-balanced Shockley pair with strong LO & principal-branch \(W_0\) approximation, valid when one diode carries nearly all the current & 76.2\% of the samples meet the \(\epsilon=0.01\) limit on the neglected current.  Over these samples, the NRMSE relative to the full two-diode solution is 0.0444\% \\
Thermionic single- and double-balanced mixers & Cardano solution (single); radical equation with a sextic residual (double) & maximum voltage differences of \(2.53\times10^{-14}\) and \(2.55\times10^{-14}\) V; normalized sextic residual of \(3.02\times10^{-15}\) \\
JFET/MESFET single- and double-balanced mixers & square-root solution (single); solution with one conducting diagonal (double) & maximum voltage differences of \(6.3\times10^{-15}\) and \(4.7\times10^{-15}\) V \\
Vacuum-triode single- and double-balanced mixers & Cardano solution (single); radical equation with a sextic residual (double) & maximum voltage differences of \(8.7\times10^{-14}\) and \(6.3\times10^{-14}\) V; normalized sextic residual of \(5.3\times10^{-15}\) \\
MOS mixers with a floating IF or a fixed center-tap bias & quartic relation with all four MOSFETs in the triode region; one-variable equation that keeps the center-tap bias & LTspice NRMSE of 0.000212\% and 0.000668\%.  Setting \(\Delta_B=0\) makes the predicted waveform differ from the correct biased equation by 32.95\% NRMSE \\
Weak-inversion MOS mixers with a fixed center tap or a floating IF & inverse of \(\sinh x+\rho x\) (fixed center tap); real \(r\)-Lambert solution (floating IF) & LTspice NRMSE of 0.000222\% and 0.005332\%.  The analytical and direct current-balance solutions agree to within rounding error \\
Near-ideal BJT mixers with emitter degeneration & successive real \(r\)-Lambert solutions for the RF and LO pairs & single- and double-balanced NRMSE of 0.002989\% and 0.003731\%.  The \(r\)-Lambert solution and the nested circuit equations agree to within \(1.4\times10^{-15}\) V \\
\bottomrule
\end{tabularx}
\end{table*}

The different circuit models answer different questions.  Circuits built
from behavioral current sources implement exactly the device laws used in
the derivations, so their agreement with the analysis confirms the algebra,
the circuit connections, and the waveform processing.  Separate runs with
LTspice's built-in JFET, MOSFET, and BJT elements, using deliberately
simplified parameters, check the device polarities and connections.  The
detailed BFU760F model, simulated at low frequency, serves as one example of
a commercial HBT.  Finally, some simulations replace the ideal transformer
voltage ratios with a set of near-ideal coupled inductors, to test the
ideal-transformer assumption.

All of these results come from simulation, and the paper reports no hardware
measurements.  The elliptic-integral result for the active MOS pair describes
a differential pair rather than a complete mixer, so it is not among the
twelve complete-mixer comparisons.
 \section{Two Hand-Calculation Design Examples}
\label{sec:design-examples-paper}

\begin{figure*}[!t]
 \centering
 \includegraphics[width=0.94\textwidth]{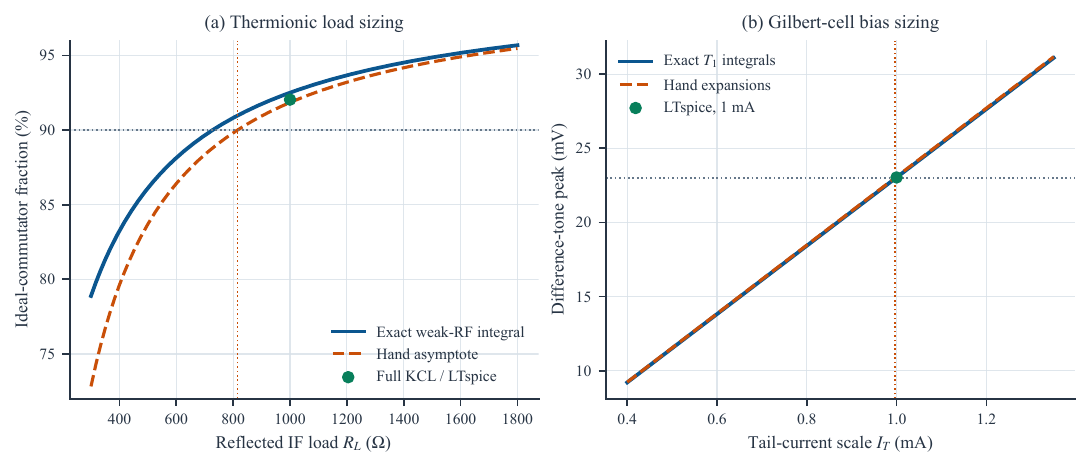}
 \caption{Hand sizing with the analytical formulas.  (a) Thermionic ring.
 The Gamma-function asymptote (dashed) reaches the 90\% target at
 \(\SI{815}{\ohm}\), a conservative choice compared with the exact weak-RF
 integral (solid).  The green point is the full KCL/LTspice result at
 \(\SI{1}{\kilo\ohm}\).  (b) Gilbert cell.  The small- and large-drive
 expansions of \(T_1\) (dashed) are almost indistinguishable from the exact
 integrals (solid) and give a tail-current scale of about
 \(\SI{1}{\milli\ampere}\) for the \(\SI{23}{\milli\volt}\) target.  The
 green point is the LTspice result at \(\SI{1}{\milli\ampere}\).}
 \label{fig:design-examples-paper}
\end{figure*}

The closed forms are more than a classification exercise.  They let a
designer work backward from a specification before opening a simulator.  The
two examples below use the same parameters and models as the validation
suite, so the hand estimate, the full equation, and LTspice can be compared
directly, with nothing refitted.  \Cref{fig:design-examples-paper}
summarizes both calculations.

\subsection{Sizing the IF load of a thermionic ring}

Consider the double-balanced Child--Langmuir ring with
\(\mathcal P=\SI{2}{\milli\ampere\per\volt\tothe{3/2}}\), \(V_c=0\),
\(\widehat\ell=\SI{6}{\volt}\), and \(\widehat r=\SI{1}{\volt}\).  The goal
is to choose the reflected IF load so that the first sideband reaches at
least 90\% of its ideal-commutator value.  With
\(\zeta=3\mathcal PR_L\sqrt{\widehat\ell}\) as in
\eqref{eq:thermionic-G1-paper}, expanding the switching integral for strong
conduction (\(\zeta\gg1\)) gives
\begin{equation}
 \begin{aligned}
 \frac{G_1}{4/\pi}&\simeq1-\frac{C_\Gamma}{\zeta},\\
 C_\Gamma&=\int_0^{\pi/2}\sqrt{\cos\theta}\,\dif\theta\\
 &=\frac{\sqrt\pi\,\Gamma(3/4)}{2\Gamma(5/4)}=1.19814.
 \end{aligned}
 \label{eq:thermionic-hand-paper}
\end{equation}
Requiring the right side to be at least \(\eta_{\rm min}=0.90\) gives the
one-line sizing rule
\begin{equation}
 R_L\gtrsim
 \frac{C_\Gamma}{3\mathcal P\sqrt{\widehat\ell}(1-\eta_{\rm min})}
 =\SI{815}{\ohm}.
 \label{eq:thermionic-load-size-paper}
\end{equation}
A designer can therefore pick \(R_L=\SI{1}{\kilo\ohm}\) directly, without
sweeping the load.  For this choice the ideal sideband is
\(2\widehat r/\pi=\SI{0.636620}{\volt}\), and the hand formula predicts
\(\SI{0.584721}{\volt}\).  The full nonlinear equation gives
\(\SI{0.585952}{\volt}\), and LTspice gives \(\SI{0.585950}{\volt}\), so the
hand estimate is 0.210\% below the simulation.  Part of this close agreement
is fortunate.  The exact weak-RF integral alone gives
\(\SI{0.588908}{\volt}\), so the error of the asymptotic expansion and the
finite-RF correction partly cancel.  The rule is a first-pass estimate for
strong conduction and weak RF; it does not check the ratings of a particular
tube or transformer.

\subsection{Sizing the tail current of a Gilbert cell}

Now suppose the double-balanced BJT cell must deliver a difference tone of
\(\SI{23}{\milli\volt}\) peak with
\(\widehat V_{\rm RF}=\SI{20}{\milli\volt}\),
\(\widehat V_{\rm LO}=\SI{200}{\milli\volt}\),
\(R_\Delta=\SI{200}{\ohm}\), unit turns ratios, and
\(U_E=\SI{25.8649}{\milli\volt}\) at \(27\,^{\circ}\mathrm C\).  The
normalized drives are then
\begin{equation}
 a_R=0.386624,\qquad a_L=3.86624.
\end{equation}
The RF drive is small and the LO drive is large.  The small-drive series in
\eqref{eq:bjt-T1-limits-paper}, and the large-drive limit refined by its
leading correction, therefore give calculator-level estimates:
\begin{align}
 T_1(a_R)&\simeq a_R-a_R^3/4+a_R^5/12=0.372896,\\
 T_1(a_L)&\simeq\frac4\pi\left(1-\frac{\pi^2}{24a_L^2}
 +O(a_L^{-4})\right)=1.238211.
\end{align}
Solving \eqref{eq:bjt-sidebands-paper} for the current gives
\begin{equation}
 I_T\simeq\frac{4\widehat V_{\rm IF}}
 {R_\Delta T_1(a_R)T_1(a_L)}=\SI{0.9963}{\milli\ampere},
 \label{eq:bjt-tail-size-paper}
\end{equation}
so \(I_T=\SI{1}{\milli\ampere}\) is the obvious choice.  Here \(I_T\) is the
available collector current.  With finite current gain, the emitter tail
source must supply approximately \(I_T/(\alpha_R\alpha_L)\).

With \(I_T=\SI{1}{\milli\ampere}\), the hand formulas predict
\(\SI{23.0862}{\milli\volt}\).  Evaluating \eqref{eq:bjt-T1-paper} by
numerical quadrature gives \(T_1(a_R)=0.37285975\) and
\(T_1(a_L)=1.23543029\), and hence
\(\widehat V_{\rm IF}=\SI{23.032111}{\milli\volt}\), while LTspice with native
BJTs returns \(\SI{23.032086}{\milli\volt}\).  The hand-sized result is
therefore within 0.235\% of the simulation.  The BFU760F vendor model gives
\(\SI{22.7561}{\milli\volt}\), a reminder that compact-device effects call
for some design margin and a final simulation with the actual device.
 \section{Discussion and Design Use}
\label{sec:discussion-paper}

Taken together, the results suggest a practical analytical workflow.
\begin{enumerate}
  \item Refer the source voltages and loads through the actual winding
  ratios, and use \eqref{eq:primary-transfer-paper} to check that the
  magnetizing inductance is large enough.
  \item Choose a device law whose range of validity covers the intended bias
  and signal swing.
  \item Specify how the IF load is connected and how its common-mode voltage
  is set.  A center tap held at a fixed voltage, separate resistors from the
  two IF endpoints to a reference, and an isolated floating winding lead to
  different circuit equations.
  \item Solve the monotone KCL equation, or its closed-form reduction where
  one is valid, and reject any candidate that violates the inequalities
  defining its conduction state.
  \item Project the result onto the IF, and pass the estimate on to
  compact-model SPICE, harmonic-balance, electromagnetic, and tolerance
  analyses.
\end{enumerate}

The same steps explain why particular special functions appear.  An
exponential current combined with a single linear load term gives the
ordinary \(W_0\).  Resistive feedback around an exponential pair gives the
real \(r\)-Lambert function.  A three-halves-power law becomes a cubic after
a square-root substitution.  Coupled square laws under a floating constraint
produce a quartic.  Keeping both \(e^x\) and \(e^{-x}\) together with a
linear term gives the inverse of \(\sinh x+\rho x\), which has no standard
name.  Bessel, Gamma/Beta, and elliptic functions enter only later, through
Fourier integration.  The device type alone therefore does not determine the
mathematics; the load and the IF connection matter just as much.

Several limitations should be kept in mind when interpreting these results.
Most of the headline comparisons test the equations against behavioral
implementations of the same memoryless device law.  They therefore verify the
algebra and the topology, not the accuracy of any commercial device model.
The native JFET and BJT runs and the vendor HBT model are useful stress
tests, but the HBT was simulated only at low frequency, and we make no claim
about its microwave performance.  The passive MOS law assumes a body held at
a fixed potential, independent of the channel terminals, and it excludes
body-diode conduction.  The triode crossbar is an actively biased circuit,
not a passive substitution.  Noise figure, compression, intercept points,
thermal behavior, and port isolation all depend on parasitics and mismatch
that are outside the scope of this paper.  Most importantly, the paper
contains no hardware measurements.  It should be judged as nonlinear circuit
theory supported by reproducible SPICE checks.

Finally, we do not claim to have invented balanced mixers, the Lambert \(W\)
function, Cardano's formula, or the Gilbert cell.  What the paper contributes
is a systematic account of which analytical solution applies to each
inductively coupled circuit, for a stated load and bias connection, and of
the conditions under which that solution is valid.
 \section{Conclusion}
\label{sec:conclusion-paper}

We have derived loaded analytical equations for inductively coupled single-
and double-balanced mixers built from six families of device laws.  The
common framework keeps three steps separate: the finite-inductance drive, the
nonlinear circuit equation under ideal transformer ratios, and the Fourier
projection onto the IF.  This separation makes it clear when a formula
derived for one IF load or common-mode connection can, and cannot, be reused
for another.

The central result is a map from physical assumptions to mathematical
structure.  The Shockley and weak-inversion exponential laws lead to the
ordinary \(W_0\), but only under their stated conditions.  Floating
weak-inversion MOS bridges and degenerated bipolar cells lead to the real
\(r\)-Lambert function.  Three-halves-power laws give Cardano cubics and
structured sextics, and floating square-law MOS bridges give Ferrari
quartics.  Spectral projection brings in Gamma/Beta, modified Bessel, and
elliptic functions.  The depletion-FET circuits remain elementary
square-root problems, and several of the complete loaded networks are best
described simply as the unique root of a monotone scalar equation.

The 37-case LTspice suite confirms all of the principal equations within
their stated conditions of validity.  The two sizing examples show the
practical value of the approach.  A load resistance or bias current can be
chosen without a parameter sweep, and the resulting IF amplitude agrees with
a simulation of the same device law to within a few tenths of a percent at
the operating points shown.  Future work should add measured transformer and
device data, conversion loss, noise, compression, and tolerance statistics,
so that the analytical front end can be tested as a complete hardware design
method.
 
\section*{Data availability}
The LTspice netlists, Python analysis programs, and simulated waveform data
that support the findings of this study are openly available on Zenodo at
\ifx\DataDOI\empty
  \textcolor{red}{\textbf{[Zenodo DOI]}}\else
  \href{https://doi.org/\DataDOI}{doi:\DataDOI}\fi.
The NXP BFU760F transistor model used in two supporting simulations is
available from the manufacturer
(\url{https://www.nxp.com/downloads/en/spice-model/BFU760F.prm}) and is not
redistributed.

\bibliographystyle{IEEEtran}

\appendix
\section{Stable Cardano Evaluation}
\label{app:stable-cardano}

The nonnegative root of \(w^3+w^2=\eta\), needed in the one-tube cases of
both the thermionic diode and the vacuum triode, can be computed without
complex intermediate values:
\begin{equation}
 w=\begin{cases}
 \displaystyle\frac23\cos\!\left[\frac13\arccos\!\left(\frac{27\eta}{2}-1\right)\right]-\frac13,
 &0\le\eta\le4/27,\\[3mm]
 \displaystyle\frac23\cosh\!\left[\frac13\arcosh\!\left(\frac{27\eta}{2}-1\right)\right]-\frac13,
 &\eta\ge4/27.
 \end{cases}
\end{equation}
For very small \(\eta\), the series
\(w=\sqrt\eta-\eta/2+5\eta^{3/2}/8+O(\eta^2)\) avoids the loss of precision
caused by subtracting nearly equal quantities.

\section{Global Single-MOS Square-Law Branch}
\label{app:mos-global}

For the single-balanced square-law circuit \eqref{eq:mos-single-square-paper},
define
\begin{align}
 \kappa_s&=\beta R_L,\qquad q_H=U+|\ell|,
 \qquad q_L=U-|\ell|,\notag\\
 \Delta&=2|\ell|,\qquad
 D_s=\tfrac12[\pospart{u_+-r}^2+\pospart{u_-+r}^2],\notag\\
 G_H&=q_H+\kappa_sD_s,\notag\\
 G_L&=q_L+\kappa_sD_s-\tfrac{\kappa_s}{2}\Delta^2.
\end{align}
We also define \(E=U+\kappa_s(D_s-\ell^2)\).  Enumerating all possible
output-side conduction states gives
\begin{equation}
 v=\begin{cases}
 -\kappa_sD_s,&G_H\le0,\\[1mm]
 \displaystyle q_H-\frac{2G_H}{1+\sqrt{1+2\kappa_sG_H}},
 &G_H>0,\ G_L\le0,\\[3mm]
 \displaystyle U-\frac{2E}{1+\sqrt{1+4\kappa_sE}},&G_L>0.
 \end{cases}
\end{equation}
The square roots are written in rationalized form so that no precision is
lost near the transitions between conduction states.

\section{Full Coupled-Inductor Descriptor Form}
\label{app:dae-paper}

Let \(\bm i_w\) collect the winding currents and \(\bm v_n\) the
nonlinear-node voltages, and let \(\bm L\) be the matrix of self and mutual
inductances, \(\bm R_w\) the winding-resistance matrix, and \(\bm A\) the
winding incidence matrix.  The complete circuit can then be written compactly
as
\begin{align}
 \bm L\dot{\bm i}_w+\bm R_w\bm i_w-\bm A^{\mathsf T}\bm v_n
 &=\bm B\bm v_s,\notag\\
 \bm A\bm i_w+\bm j_{\rm nl}(\bm v_n,\bm q,t)
 +\bm G\bm v_n+\frac{\dif}{\dif t}\bm q(\bm v_n)&=\bm0,
 \label{eq:full-dae-paper}
\end{align}
where \(\bm B\) connects the source voltages \(\bm v_s\), \(\bm j_{\rm nl}\)
collects the nonlinear device currents, \(\bm G\) is the nodal conductance
matrix, and \(\bm q\) is the stored node charge.  With ideal coupling,
\(\bm L\) can become singular, so \eqref{eq:full-dae-paper} is in general a
descriptor (differential-algebraic) system.  The scalar equations in the body
of the paper follow when the transformer voltage ratios are imposed
algebraically and charge storage, leakage, and winding loss are removed.

\section{Validation Procedure and Scope}
\label{app:reproducibility-paper}

The accompanying validation material assigns each of the 37 LTspice netlists
to a specific equation or check before any simulation is run.  A driver
script then runs the netlists one at a time, confirms that each run writes
fresh raw-data and log files, executes the seven analysis programs, and
collects the resulting metrics.  Two items are deliberately excluded from the
twelve primary complete-mixer comparisons: an archived worksheet for an
earlier, inverse-sinh reduced model of the double ring, and the active-MOS
differential pair, which is a sub-circuit rather than a complete mixer.
Earlier development netlists are not part of the reported results, and no
hardware data are used.
 
\end{document}